\documentclass[aps,prb,twocolumn,showpacs,10pt]{revtex4-1}

\usepackage{latexsym}
\usepackage{graphicx}
\usepackage{times,psfrag,subfigure}
\usepackage{amsmath}
\usepackage{dcolumn}
\usepackage{color}
\usepackage{latexsym,amsmath,amssymb,bm,euscript}
\usepackage{amssymb,amsmath,amsfonts,hyperref}
\usepackage{wasysym}
\usepackage{latexsym}
\usepackage{subfigure}

\newcommand{\mc}{\mathcal}

\newcommand{\bs}{\boldsymbol}

\begin{document}
\title{Exotic phases induced by strong spin-orbit coupling in ordered double perovskites}

\date{\today}

\author{Gang Chen}
\affiliation{Physics Department, 
University of Colorado, Boulder, CO 80309}
\affiliation{Physics Department, 
University of California, Santa Barbara, CA 93106}
\author{Rodrigo Pereira}
\affiliation{Instituto de F\'{\i i}sica de S\~ao Carlos, Universidade de S\~ao Paulo, C.P. 369, 
S\~ao Carlos, SP.  13566-970, Brazil}
\affiliation{Kavli Institute for Theoretical Physics, 
University of California, Santa Barbara, CA 93106}
\author{Leon Balents}
\affiliation{Kavli Institute for Theoretical Physics,
University of California, Santa Barbara, CA 93106}

\begin{abstract}
  We construct and analyze a microscopic model for insulating rock
  salt ordered double perovskites, with the chemical formula
  A$_2$BB'O$_6$, where the B' atom has a 4d$^1$ or 5d$^1$ electronic
  configuration and forms a face centered cubic (fcc) lattice.  The
  combination of the triply-degenerate $t_{2g}$ orbital and strong
  spin-orbit coupling forms local quadruplets with an effective spin
  moment $j=3/2$.  Moreover, due to strongly orbital-dependent
  exchange, the effective spins have substantial biquadratic and
  bicubic interactions (fourth and sixth order in the spins,
  respectively).  This leads, at the mean field level, to three main
  phases: an unusual antiferromagnet with dominant octupolar order, a
  ferromagnetic phase with magnetization along the $[110]$ direction,
  and a non-magnetic but quadrupolar ordered phase, which is
  stabilized by thermal fluctuations and intermediate temperatures.
  All these phases have a two sublattice structure described by the
  ordering wavevector ${\bs Q} =2\pi (001)$.  We consider quantum
  fluctuations and argue that in the regime of dominant
  antiferromagnetic exchange, a non-magnetic valence bond solid or
  quantum spin liquid state may be favored instead.  Candidate quantum
  spin liquid states and their basic properties are described. We also
  address the effect of single-site anisotropy driven by lattice
  distortions.  Existing and possible future experiments 
  are discussed in light of these results.
\end{abstract}
\date{\today}

\pacs{71.70.Ej,71.70.Gm,75.10.-b}


\maketitle

\section{Introduction}
\label{sec:sec1}

In magnetic Mott insulators with quenched orbital degrees of freedom,
weak spin-orbit coupling (SOC) only leads to a small correction to the
usual spin exchange Hamiltonian in the form of single-site anisotropy
and Dzyaloshinskii-Moriya
interactions.\cite{PhysRev12091,Dzyaloshinsky1958241} In the presence of
strong SOC, however, a completely different physical picture emerges, in
which spin itself is not a good quantum number, and magnetic anisotropy
is usually large.  Generally, strong SOC is common in the Lanthanides,
in which the relevant 4f-electrons are very tightly bound to the
nucleus.  The tight binding shields the electrons from crystal fields,
which tend to split the orbital degeneracies involved in SOC, and
moreover reduces exchange, which also competes with SOC.  

While more rare, strong SOC is becoming an increasing focus in
d-electron systems, in which electrons are more delocalized than in the
Lanthanides, and more diverse phenomena can be expected.  For instance,
strong SOC can be expected in 5d transition metal compounds, which have
large intrinsic atomic SOC due to their high atomic weight. In this
category, many Ir-based magnets have been studied
recently\cite{PhysRevB.78.094403,PhysRevLett.102.017205}.  Lighter
transition metals may also exhibit strong SOC if competing effects such
as crystal fields and exchange are suppressed, e.g.  by choosing crystal
structures with high-symmetry and well-separated magnetic ions,
respectively.  An example of this type is the ``spin-orbital liquid''
state observed in the Fe-based spinel
FeSc$_2$S$_4$,\cite{loidl:nsnmr,fritsch:prl04,loidl:ns} which is
believed to be driven by SOC. \cite{PhysRevLett.102.096406,PhysRevB.80.224409,PhysRevB.82.041105}

In this paper, we consider the case of insulating magnetic ordered
double perovskites.  Structurally, ordered double perovskites (with
the chemical formula A$_2$BB'O$_6$) are derived from the usual
perovskites ABO$_3$ by selectively replacing half the B ions with
another species, denoted B'.  We focus on the case in which the B
ions are non-magnetic and the B' ones are magnetic.  Because of the
difference in the valence charges and ionic radius between B and
B' ions, the magnetic B' ions form an fcc lattice structure with a
lattice constant double of the original cubic one.  Many ordered
double perovskites incorporate strong intrinsic SOC, as B' ions are
commonly 4d and 5d transition metals.  Moreover, the large B'-B'
separation weakens exchange, similarly to FeSc$_2$S$_4$.  Here, we
construct an appropriate microscopic model for the most quantum of
these materials (a list may be found in Table~\ref{tab:Tab1}), in
which the magnetic ion contains a single unpaired $S=1/2$ spin. 

The physics is strongly influenced by the combination of the orbital
degeneracy of the $t_{2g}$ multiplet, which acts as an effective
$\ell=1$ orbital angular moment.  Due to strong SOC, this combines
with the $S=1/2$ spin to induce an effective total angular momentum
$j=3/2$ description of the system.  Moreover, due to the
orbitally-dependent exchange, the interaction of these $j=3/2$
contains large biquadratic (fourth order in spin operators) and
triquadratic (sixth order in spin operators) interactions.  These
support exotic phases not easily found in systems with dominant
bilinear spin exchange.

Analysis of the microscopic model shows that the strong SOC
enhances quantum fluctuations and leads to several interesting phases:
(1) an unconventional antiferromagnet (denoted AFM) in which the
magnetic {\sl octupole} and {\sl quadrupole} moments rather than the
dipole moment are dominant, (2) an unusual non-collinear ferromagnet
(denoted FM110) with a doubled unit cell and magnetization along the
$[110]$ axis, (3) a (biaxial) ``spin nematic'' phase with
quadrupolar order but unbroken time reversal symmetry and, more
speculatively, (4) a possible quantum spin liquid (QSL) phase.  Phases
(1), (2), and (4) are low temperature phases and persist as ground
states, while the spin nematic, phase (3), occurs in a broad
intermediate temperature range below the paramagnetic state but above
any magnetic ordering temperature.

\begin{figure}[htp]
\includegraphics[width=8cm]{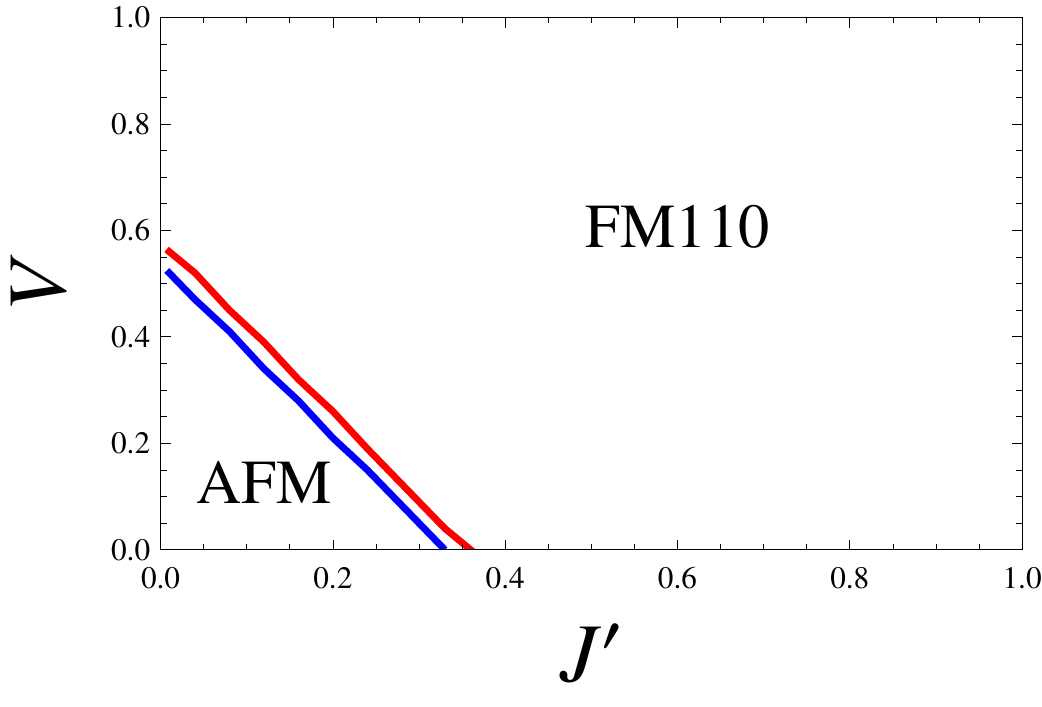}
\caption{(Color online) Mean field $T=0$ phase diagram for the model
  Hamiltonian in Eq.~\eqref{model}. AFM denotes the
  ``antiferromagnetic'' ground state given by Eq.~\eqref{eq:state_aniz1}
  and Eq.~\eqref{eq:state_aniz2}, FM110 denotes the ferromagnetic ground
  state with an easy axis oriented along $[ 110]$, given by
  Eq.~\eqref{eq:stateFM1}.  The FM100 state, which is ferromagnetic with
  easy axis along $[ 100]$ appears in the narrow band
  between the two phase phase boundaries.  In the figure, $J=1$.}
\label{fig:phase_diagram}
\end{figure}

States with magnetic multipole order are more often observed in
f-electron systems where crystal field effects are less important than
SOC.\cite{RevModPhys.81.807} As a consequence, the atomic wavefunctions
are total angular momentum eigenstates, in which the spin and orbital
degrees of freedom are highly entangled.  This leads to highly
non-Heisenberg exchange between the local moments, which is described by
interaction of higher magnetic multipole operators.  Such interactions
may drive multipolar order, as suggested for instance in
URu$_2$Si$_2$.\cite{Kotliar} Recently this has been suggested to also
occur in d electron systems with unquenched orbital degeneracy and
sufficient SOC.\cite{PhysRevLett.103.067205}\   We find a similar
mechanism at work in the AFM phase.  

A ferromagnetic state is not in itself unusual, though such is
relatively uncommon in insulators.  However, cubic ferromagnets with an
easy axis oriented along the $[110]$ direction is quite
uncommon.  This can be understood from the Landau theory for a
ferromagnet: the usual fourth order cubic anisotropy term favors either
$[100]$ or $[111]$ orientation, depending
upon its sign, but never $[110]$.  To obtain
a $[110]$ easy axis, one requires sixth order or higher
terms to be substantial, making this rare indeed.  Remarkably, such
$[110]$ anisotropy has been observed in experiments on
Ba$_2$NaOsO$_6$.\cite{Erickson} 

Both the above states, when heated above their magnetic ordering
temperatures, allow on symmetry grounds for an intermediate phase which
is time-reversal symmetric but with quadrupolar order -- the spin
nematic.  Applying the mean field theory at $T>0$, we indeed find such a
phase in a broad range of parameter space.  While spin nematic states
have been suggested previously in
NiGa$_2$S$_4$\cite{stoudenmire:214436,läuchli2006quadrupolar,tsunetsugu-06,senthil-06,nakatsuji-05},
it has not been established in that material.  The mechanism for
quadrupolar order here is much more transparent and robust than in that
case.

The above three phases, while somewhat unconventional, may be obtained
within a mean-field analysis.  A QSL state, however, cannot be
described by any mean field theory, and is considerably more exotic.  
The search for a QSL, which is a state in which
quantum fluctuations prevent spins from ordering even at zero
temperature, is a long-standing problem in fundamental
physics.\cite{balents:nat} Since the possibility of a QSL was suggested
by Anderson in the early 1970s,\cite{Anderson1973153} this has been an
active area for theory and experiment.  Despite the current maturity of
the theory for QSL,\cite{PhysRevB.65.165113} the experimental
confirmation of the existence of such an exotic phase is still elusive.
Very commonly geometrical frustration is thought to be a driving
mechanism for QSL formation, and consequently most research (both
theoretically and experimentally) has been devoted to systems of this
type, such as triangular,\cite{PhysRevB.73.155115}
kagome,\cite{PhysRevB.75.184406}
hyperkagome\cite{PhysRevB.78.094403,PhysRevLett.99.137207,PhysRevLett.101.197202}
and pyrochlore lattices.\cite{PhysRevB.69.064404} 

Here we suggest a
different route, in which quantum fluctuations are enhanced primarily by
strong SOC, rather than geometrical frustration.  In fact, the magnetic
ions in ordered double perovskites reside on a face centered cubic (fcc)
sublattice, which can be viewed as edge-sharing tetrahedra, and is
somewhat geometrically frustrated.  Without strong SOC, however, this
frustration is weak, and indeed the classical Heisenberg antiferromagnet
on the fcc lattice is known to magnetically order into a state with the
ordering wavevector $2\pi(001)$.\cite{henley} The tendency of the simple
fcc antiferromagnet to order may be partially attributed to its large
coordination number ($z=12$), which leads to mean-field like behavior.
By contrast, strong SOC induces effective exchange interactions very
different from Heisenberg type, with strong directional dependence that
may make a QSL more favorable.  To make this suggestion more concrete,
we propose a natural wavefunction for a QSL in our model, and discuss
the physical properties of such a state.   

We now outline the main results of the paper, and how they are presented
in the following sections.  In Sec.~\ref{sec:sec2}, we show that strong
SOC leads to an effective $j=3/2$ local moment on each B' site. We
write down a model Hamiltonian which includes three interactions:
nearest neighbor (NN) antiferromagnetic (AFM) exchange, $J$, NN
ferromagnetic (FM) exchange, $J'$, and electric quadrupolar interaction,
$V$.  These interactions are all {\sl projected} down to the effective
$j=3/2$ manifold, which induces many terms beyond the usual quadratic
exchange.  Indeed, because of the four-dimensional basis of spin-$3/2$
states, the resulting Hamiltonian can be thought of as an anisotropic
$\Gamma$ matrix model.\cite{PhysRevB.69.235206} We then discuss the
symmetry properties of the projected Hamiltonian. Surprisingly, we find
that, in the limit of vanishing FM exchange, the Hamiltonian has a
``hidden'' global SU$(2)$ symmetry despite its complicated appearance.

In Sec.~\ref{sec:sec3}, we consider the mean field ground states of the
model, characterized by local (single-site) order parameters.  In
Sec.~\ref{sec:sec31} we begin by considering the more accessible limit
in which strong uniaxial single-site anisotropy (due e.g. to a
tetragonal distortion of the crystal) lifts the ``orbital'' four-fold
degeneracy of $j=3/2$ quadruplets down to easy-axis or easy-plane
Kramer's doublets.  In these limits, the effective Hamiltonian in the
reduced phase space is mapped onto that of an XXZ antiferromagnet which
can be understood even without mean field theory.  Next, in
Sec.~\ref{sec:sec32}, we carry out $T=0$ mean field theory for the case
of cubic symmetry.  Here we find the AFM state and two ferromagnetic
states (the FM110 state and another state with a $[ 100]$
easy axis).  The $T=0$ mean field phase diagram is shown in
Fig.~\ref{fig:phase_diagram}.  Finally, having described the situations
with strong and vanishing single-site anisotropy, we determine in
Sec.~\ref{sec:interm-anis} the mean-field phase diagram for intermediate
anisotropy.

In Sec.~\ref{sec:sec4} we identify the multipolar order parameters of
the three ordered phases and analyze the $T>0$ behavior by mean-field
theory.  Here we find the quadrupolar phase, and discuss several phase
transitions which occur.  We also discuss the behavior of the magnetic
susceptibility in different parameter regimes.

In Sec.~\ref{sec:sec5}, we consider quantum effects beyond the mean
field theory.  First, we carry out a spin wave calculation, which
determines the collective mode structure, and also shows that in the
regime where nearest-neighbor antiferromagnetic exchange is dominant
(small $J'$ and $V$) quantum fluctuations are large and may destabilize
the ordered AFM phase.  Therefore, we consider possible non-magnetic
ground states, both of valence bond solid and quantum spin liquid (QSL)
type.  We formulate a slave-fermion theory with four-component spin
$s=3/2$ fermions, in such a way that at mean field level the hidden
SU$(2)$ symmetry is preserved and the correct ground state, the analog
of a singlet in the usual Heisenberg model, is obtained for a single
pair of nearest-neighbor sites.  The corresponding mean field theory
naturally includes the intrinsic spatial anisotropy of the strong SOC
limit. We analyze two different mean field ans\"atze, with zero and
$\pi$-flux. In both cases the mean field Hamiltonian respects all the
symmetries of the original spin Hamiltonian. The $\pi$-flux state is
found to have lower mean field energy. For both states, the spinons are
at quarter-filling, leading to a spinon Fermi sea.  There is no Fermi
surface nesting and we expect that this spinon Fermi surface should be
stable against weak perturbations.  Predictions based on the picture of
spinon Fermi surface are made.

Finally in Sec.~\ref{sec:sec6}, we compare our theoretical prediction
with current experimental findings and suggest further directions for
theory and experiment.

\section{Model and Symmetry}
\label{sec:sec2}


\subsection{Spin-orbit interaction and hybridization of atomic orbitals}

The magnetic ions B' (Os$^{7+}$, Re$^{6+}$, Mo$^{5+}$) found in the
ordered double perovskites in Table~\ref{tab:Tab1} all have one electron
in the triply degenerate $t_{2g}$ multiplet. The atomic spin-orbit
interaction projected down to the $t_{2g}$ triplet is written as
\begin{equation}
{\mc H}_{\text{so}} = - \lambda \; {\bm l}\cdot {\bs S}
\;,
\end{equation}
in which the total angular momentum quantum numbers of these operators
are $l=1,S=1/2$. The effective orbital angular momentum ${\bs l}$ comes
from the projection of orbital angular momentum ${\bs L}$ onto the
$t_{2g}$ triplets,
\begin{equation}
\mathcal{P}_{t_{2g}} {\bs L}\mathcal{P}_{t_{2g}} =  - {\bs l}
\;.
\end{equation}
Here $\mathcal{P}_{t_{2g}} = \sum_{a=yz,xz,xy} |a\rangle \langle a|$ is
the projection operator to the $t_{2g}$ manifold.  The eigenstates of
$l^z$ with eigenvalues $m=0,\pm 1$ and $S^z$ with eigenvalues
$\sigma=\pm 1/2 \equiv \uparrow,\downarrow$, written in terms of the
usual $t_{2g}$ states are \begin{equation}
|0,\sigma\rangle=|d^\sigma_{xy}\rangle;\quad
|\pm1,\sigma\rangle=\frac{\mp|d^\sigma_{yz}\rangle-
  i|d^\sigma_{xz}\rangle}{\sqrt2}.  \end{equation}

This interaction favors $j=3/2$ ($\bs{j} = \bs{l}+ \bs{S}$) quadruplets over $j=1/2$ doublets by an energy separation $3\lambda/2$. In the strong spin-orbit interaction limit, the local Hilbert space is restricted to four low-lying states 
\begin{equation}
|d_\alpha\rangle=\sum_{m,\sigma}C_{m\sigma}^\alpha |m,\sigma\rangle,
\end{equation}
where $\alpha=\pm3/2,\pm1/2$ is the $j^z$ eigenvalue and
\begin{equation}
C_{m\sigma}^\alpha=\left\langle l=1,S=\frac12;m,\sigma \left| l=1,S=\frac12;j=\frac32,\alpha\right.\right\rangle
\end{equation}
is a Clebsch-Gordan coefficient.  In the materials under consideration,
$\lambda$ is indeed a very large energy scale (fraction of an eV),
justifying the strong SOC limit.  

Every operator expressed in terms of spin and orbitals must therefore be projected
into this subspace and its projection can be written in terms of $j=3/2$
angular momentum operator. For example, 
\begin{eqnarray}
{\mc P}_{\frac{3}{2}}\;\bs{S}\;{\mc P}_{\frac{3}{2}} &=& \frac{1}{3} \; \bs{j}, \\
{\mc P}_{\frac{3}{2}}\;\bs{l}\;{\mc P}_{\frac{3}{2}} &=& \frac{2}{3} \;
\bs{j} \;.  
\end{eqnarray} 
Here ${\mc P}_{\frac{3}{2}}$ is the projection operator
into the $j=3/2$ quadruplets. Furthermore, for the magnetic moment ${\bs
  M}$ for electrons in atomic $d$ orbitals, we have 
\begin{equation} {\bs M} \equiv
{\mc P}_{\frac{3}{2}} [2 \bs{ S} + (-\bs{ l})]{\mc P}_{\frac{3}{2}} = 0
\;.\label{zeromom} 
\end{equation} 
The vanishing magnetic moment is quite remarkable
and partially explains why the compounds have small magnetic moments in
comparison with spin-$\frac{1}{2}$ systems without orbital
degeneracy. 

In reality, the measured magnetic moments are nonzero because the atomic
4d or 5d orbitals strongly hybridize with p orbitals at the oxygen
sites that form an octahedron surrounding each B' site. For instance,
for Ba$_2$NaOsO$_6$, the hybridization energy is estimated to be of the
order of electron volts \cite{lee:epl2007, Erickson} and comparable to
the energy gap between Os d and O p states. For this reason, it is
more appropriate to think in terms of molecular orbitals with mixed d
and p character. For example, molecular $xy$ orbitals are written as
\begin{equation}
|D^\sigma_{i,xy}\rangle=\frac{|d^\sigma_{i,xy}\rangle+r|p^\sigma_{i,xy}\rangle}{\sqrt{1+r^2}},\label{eq:1}
\end{equation}
where $|d^\sigma_{i,xy}\rangle$ is the state corresponding to one
electron in the $xy$ orbital and spin $\sigma$ on site $i$, and
$|p^\sigma_{i,xy}\rangle$ is a linear combination (with $xy$ symmetry)
of states that have a singlet on the $ d_{xy}$ orbital and one hole on
an oxygen site
\begin{equation}
|p^\sigma_{i,xy}\rangle=\frac12\left(|p^\sigma_{i+\hat{e}_x,y}\rangle+|p^\sigma_{i+\hat{e}_y,x}\rangle+|p^\sigma_{i-\hat{e}_x,y}\rangle+|p^\sigma_{i-\hat{e}_y,x}\rangle\right),
\end{equation}
where $\hat{e}_{x,y}$ are real space vectors from the B' site to
neighboring oxygens along $x$ or $y$ directions. The mixing parameter
$r$ is of order $t_{dp}/\Delta$, where $t_{dp}$ is the hopping matrix
element between d and p orbitals and $\Delta$ is the gap to oxygen
p states.  In the limit of strong spin-orbit interaction, we must
project into four low-lying molecular orbitals which are a superposition
of the four atomic states with $j=3/2$ and p states
\begin{equation}
|D_{i,\alpha}\rangle=\sum_{m,\sigma}C_{m\sigma}^\alpha \left| D_{i,m}^\sigma\right\rangle.
\end{equation} 
While the atomic magnetic moment in Eq. (\ref{zeromom}) vanishes, there
is a nonzero contribution to the molecular $\bs{M}$ from holes in p
orbitals. After taking the projection into $j=3/2$ states, the coupling
of the molecular orbital to a magnetic field reads
\begin{equation}
\mc{H}_{Z}=-g \mu_B \, \bs{h}\cdot \bs{j},
\end{equation}
where $g=r^2/[3(1+r^2)]$ is the Land\'e factor, and $\mu_B$ is the Bohr
magneton.  

\subsection{Exchange interactions and electric quadrupolar interaction}

In the last subsection, we discussed the effect of strong spin-orbit
interaction in determining the local degrees of freedom and pointed out
that every operator must be projected into the $j=3/2$ quadruplets. In
this subsection, we introduce the interactions between the local
moments, and discuss the mechanics of the projection.

\begin{figure}[htp]
\subfigure{\includegraphics[width=3.5cm]{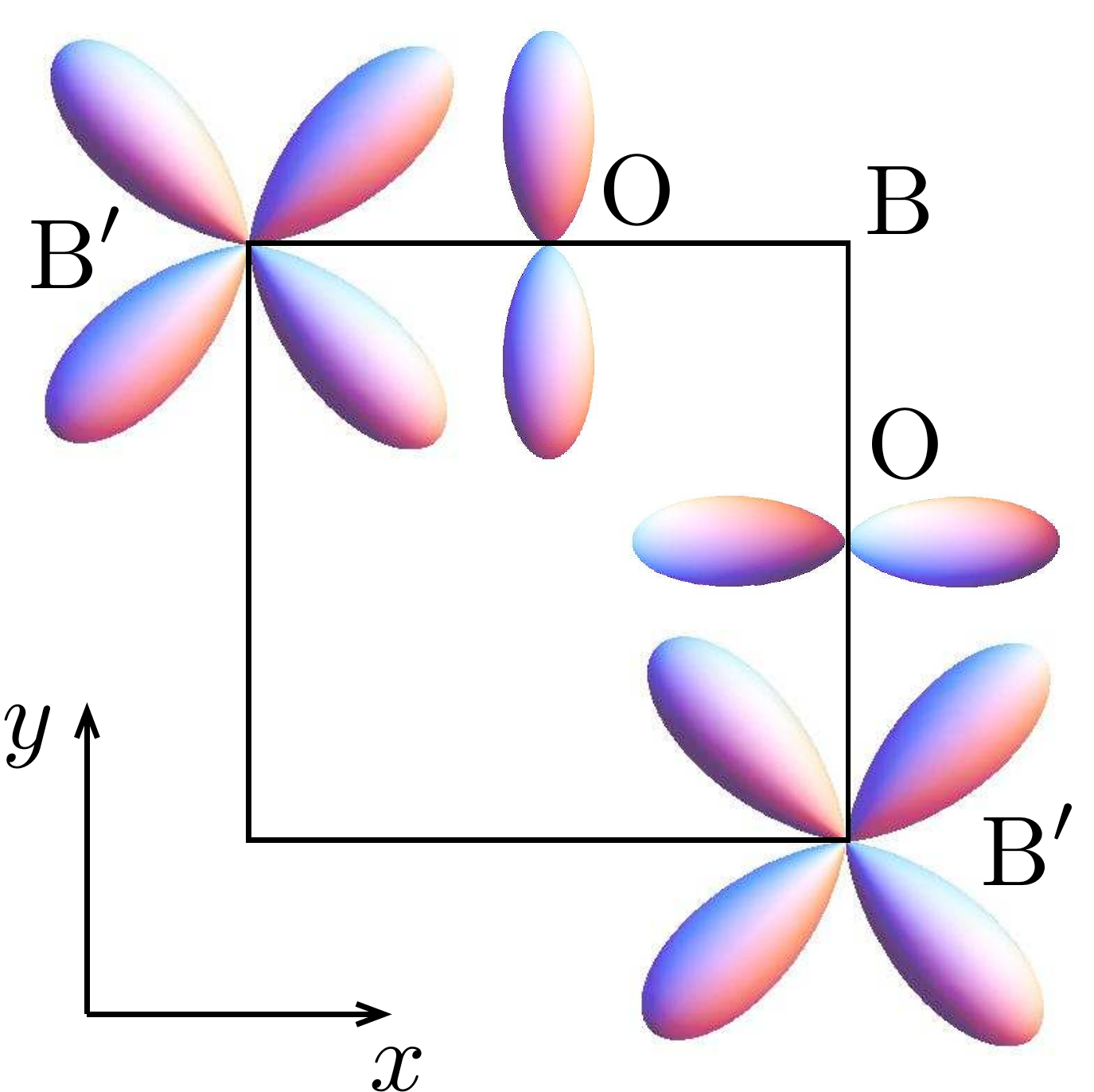}}
\subfigure{\includegraphics[width=4.5cm]{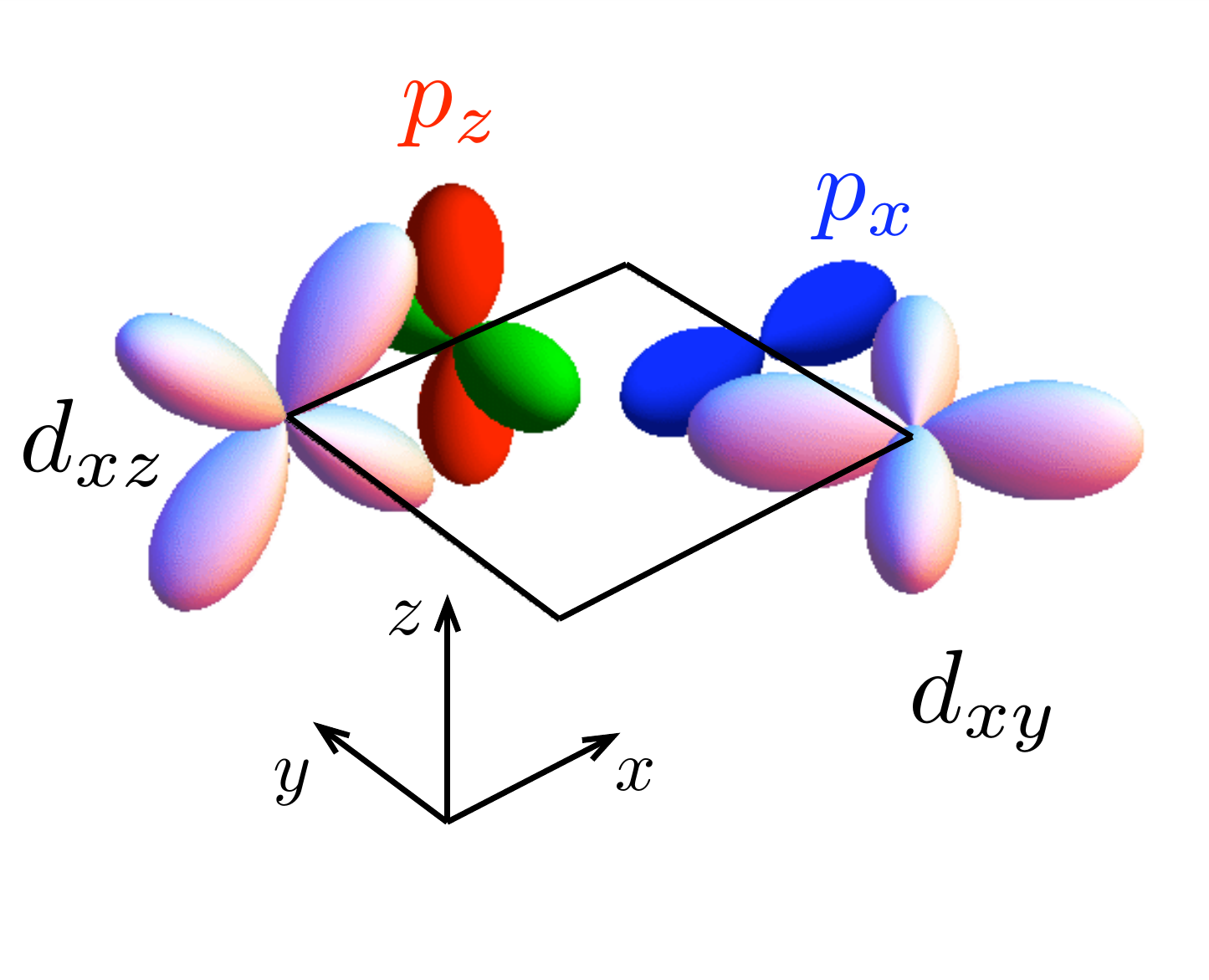}}
\caption{(Color online) Left graph: The NN AFM exchange path (B$'$-O-O-B$'$); right graph: The NN FM exchange path with
intermediate orthogonal $p$ orbitals at O sites. }
\label{fig:fig1}
\end{figure}

The first interaction to consider is nearest-neighbor antiferromagnetic
exchange, through the virtual transfer of electrons through intermediate
oxygen p orbitals.  These processes are strongly restricted by
symmetry.  For example, in XY planes, only electrons residing on
$d_{xy}$ orbitals can virtually hop to neighboring sites via $p_x$ and
$p_y$ orbitals of the intermediate oxygen sites.  The exchange path and
relevant orbitals are depicted in Fig.~\ref{fig:fig1}. Alternatively,
one can interpret this process as kinetic exchange between molecular
$D_{xy}$ orbitals, which are mixtures of the transition metal d state
and p states on the neighboring four oxygens (see Eq.~\eqref{eq:1}).
As a consequence, the antiferromagnetic exchange interaction can be
written $\mathcal{H}_{\text{ex-1}} =\mathcal{H}_{\text{ex-1}}^{\text{XY}}
+\mathcal{H}_{\text{ex-1}}^{\text{YZ}} +\mathcal{H}_{\text{ex-1}}^{\text{XZ}} $, where
\begin{equation} 
{\mathcal H}_{\text{ex-1}}^{\text{XY}} =  
J \sum_{\langle ij \rangle \in \text{XY}}\left(
\bs{S}_{i,xy} \cdot \bs{S}_{j,xy} - \frac{1}{4}\; n_{i,xy}  n_{j,xy}\right)
\;,
\end{equation}
where the sum is over nearest neighbor sites in the XY planes, and the
corresponding terms for YZ and XZ planes are obtained by the obvious
cubic permutation.  Here the operators $\bs{S}_{i,xy}$ and
$n_{i,xy}$ denote the spin residing on $xy$ orbital and orbital
occupation number at site $i$, respectively. In terms of spin and
orbital angular momentum operators acting on site $i$,
\begin{eqnarray}
\bs{S}_{i,xy}&=&\bs{S}_i[1-(l_i^z)^2],\\
n_{i,xy}&=&1-(l_i^z)^2,
\label{eq:spinnumber}
\end{eqnarray} 
Throughout this paper, we use the subindices ($i,xy$) to denote the site
and orbitals, superindex ($\mu=x,y,z$) to denote the spin component, and
capital letters (XY, XZ, YZ) to denote the planes.  With these
definitions, we note that the 
single occupancy condition at each site, which defines the Mott
insulating state, becomes
\begin{equation}
n_{i,xy}+n_{i,xz}+n_{i,yz} = 1.
\label{singleoc}
\end{equation} 
Moreover, from Eq.~\eqref{eq:spinnumber}, orbitally-resolved spins satisfy
\begin{equation}
\bs{S}_{i,xy} + \bs{S}_{i,yz}+ \bs{S}_{i,xz} = \bs{S}_i
\;.
\label{eq:spinconstraint}
\end{equation}

The second interaction is the nearest-neighbor ferromagnetic exchange
interaction. This interaction is due to the spin transfer through
orthogonal orbitals at the intermediate oxygen sites in the exchange
path, as shown in Fig.~\ref{fig:fig1}. For two sites $i,j$ in the XY
plane, this ferromagnetic exchange is written as
\begin{eqnarray}
{\mathcal H}_{\text{ex-2},ij}^{\text{XY}} &=&
- J'  \left[
\bs{S}_{i,xy} \cdot (\bs{S}_{j,yz}+\bs{S}_{j,xz})   +\langle i \leftrightarrow j\rangle  \right]
\nonumber \\
& &
 - \frac{3J'}{4}  \left[n_{i,xy} (n_{j,xz}  +n_{j,yz}) +\langle i \leftrightarrow j\rangle \right]
\;,
\end{eqnarray}
where the $xy$ orbital only interacts with $yz$ and $xz$ orbitals at neighboring sites. 
Applying the single-occupancy constraint, the nearest-neighbor ferromagnetic exchange 
interaction can be simplified, up to a constant, to
\begin{eqnarray}
{\mathcal H}_{\text{ex-2}}^{\text{XY}} &=&
- J'   \sum_{\langle ij \rangle \in \text{XY}} \left[
\bs{S}_{i,xy} \cdot (\bs{S}_{j,yz}+\bs{S}_{j,xz})   +\langle i \leftrightarrow j\rangle  \right]
\nonumber \\
& &
 + \frac{3J'}{2}  \sum_{\langle ij \rangle} n_{i,xy} n_{j,xy}  
\;.
\label{exchangeFM2}
\end{eqnarray}

Microscopically, $J'/J \sim {\mc O}(J_H/U_p) $ where $J_H$ and $U_p$ are
the Hund's coupling and Hubbard Coulomb interaction at the oxygen site,
respectively.

The third interaction is the electric quadrupole-quadrupole
interaction. The 4d or 5d electron carries an electric quadrupole
moment, and the interaction between these moments may not be negligible
because of the long spatial extent of the molecular
orbitals. Calculating the direct electrostatic energy between all
possible orbital configurations for two electrons residing in
neighboring sites in an XY plane, we obtain the
quadrupole-quadrupole interaction
\begin{eqnarray}
{\mc H}_{\text{quad},ij}^{\text{XY}}& =& V n_{i,xy}n_{j,xy}\nonumber\\
&&-\frac{V}2[n_{i,xy}(n_{j,yz}+n_{i,xz})+(i\leftrightarrow j)]\nonumber\\
&&-\frac{13V}{12}(n_{i,yz}n_{j,yz}+n_{i,xz}n_{j,xz})\nonumber\\
&&+\frac{19V}{12}(n_{i,yz}n_{j,xz}+n_{i,xz}n_{j,yz}).
\end{eqnarray}
Here $V>0$ is defined as the Coulomb repulsion between two nearest-neighbor $xy$ orbitals on XY planes. If $Q$ is the magnitude of the electric quadrupole and $a$ is the lattice constant of the fcc lattice, we have $V=9\sqrt2Q^2/a^5$.  In general, the main contribution to $Q$ comes from the charge at the oxygen sites, hence the larger the hybridization, the larger the value of $V$. Using the single-occupancy constraint in Eq. (\ref{singleoc}) and summing over sites, the quadrupole-quadrupole interaction simplifies to  
\begin{eqnarray} 
{\mc H}_{\text{quad}}^{\text{XY}} &=&  \sum_{\langle ij \rangle\in \text{XY}}\left[- \frac{4V}{3} (n_{i,xz} - n_{i,yz}) (n_{j,xz} - n_{j,yz})
 \right.\nonumber\\
&&\left.+\frac{9V}{4} n_{i,xy} n_{j,xy}
\right],
\label{quadrupole2}
\end{eqnarray}
in which we have ignored an unimportant constant.

The minimal Hamiltonian for the cubic system contains all three of these
exchange interactions in addition to the on-site SOC,
\begin{equation} 
{\mc H} = {\mc H}_{\text{ex-1}} + {\mc H}_{\text{ex-2}} + {\mc H}_{\text{quad}} + {\mc H}_{\text{so}}
\;.
\end{equation}

Since we are interested in the limit of strong spin-orbit interaction, we need to project ${\mc H}$ onto the $j=3/2$ quadruplets at every site. As an example, we write down the projection for $\bs{S}_{i,xy}$ and $n_{i,xy}$,
\begin{eqnarray}
\label{eq:projxy}
 \tilde{S}_{i,xy}^x &=&\frac{1}{4} j_i^{x} -\frac{1}{3} {j_i^z j_i^{x} j_i^z} \\
 \tilde{S}_{i,xy}^y &=& \frac{1}{4}{j_i^{y}}- \frac{1}{3}{j_i^z j_i^{y} j_i^z}\\
 \tilde{S}_{i,xy}^z &=& \frac{3}{4} j_i^z -\frac{1}{3} j_i^z j_i^{z} j_i^z \\
 \tilde{n}_{i,xy}     &=& \frac{3}{4} - \frac{1}{3}(j_i^z)^2,
\end{eqnarray}
in which, $ \tilde{\mc O} \equiv {\mc P}_{\frac{3}{2}}\; {\mc O} \; {\mc
  P}_{\frac{3}{2}} $. Spin and occupation number operators for other
orbitals can be readily generated by a cubic permutation. After the
projection, the minimal Hamiltonian reduces, up to a constant, to
\begin{equation}
\tilde{\mc H} = \tilde{\mc H}_{\text{ex-1}} + \tilde{\mc H}_{\text{ex-2}} +  \tilde{\mc H}_{\text{quad}} 
\;.\label{model}
\end{equation}

As one may notice, the projected Hamiltonian contains 4-spin and 6-spin
interactions in addition to the usual quadratic 2-spin interactions if it
is expressed in terms of the effective spin moment ${\bs j}_i$.  One can
view these multiple spin terms as the interaction between magnetic
multipoles (quadrupole and octupole) at different sites. Such multipolar
Hamiltonians are much less familiar than the usual quadratic exchange
forms, and some caution should be used.  In particular, experience with
similar models shows that such interactions can magnify quantum effects,
for instance leading to the appearance of a quadrupolar phase in the
biquadratic case\cite{läuchli2006quadrupolar}.  Hence, the na\"ive
classical approximation -- replacing $\bs{j}$'s by classical vectors --
is inadvisable, and we will proceed differently below.

\subsection{Symmetry properties of the Hamiltonian}

Before we move on to discuss the ground state of the Hamiltonian
$\tilde{\mc H}$ in Eq. (\ref{model}), we need to have some understanding
about its symmetry properties. We start from the NN AFM exchange
interaction $\tilde{\mc H}_{\text{ex-1}}$. The latter has an apparent
cubic space group symmetry. The total angular momentum
$\bs{J}=\sum_i\bs{j}_i$ is \emph{not} conserved, $[\tilde{\mc
  H}_{\text{ex-1}},\bs{J}]\neq0$. Nevertheless, $\tilde{\mc
  H}_{\text{ex-1}}$ surprisingly has a ``hidden'' SU$(2)$ symmetry. The
three generators of this global \emph{continuous} symmetry are defined
as follows,
\begin{equation}
G^{\mu}  = \sum_iG^\mu_i= \sum_i \left[\frac{7}{6} j_i^{\mu} - \frac{2}{3} (j_i^{\mu})^3 \right],
\end{equation}
with $\mu = x,y,z$.
One can readily check that these generators commute with $\tilde{\mc H}_{\text{ex-1}}$,
\begin{equation} 
[G^{\mu}, \tilde{\mc H}_{\text{ex-1}}] =0 ,
\end{equation}
and satisfy the SU$(2)$ algebra,
\begin{equation}
[G^{\mu}, G^{\nu}] = i \epsilon_{\mu\nu\lambda} G^{\lambda}
\;.
\end{equation}
In addition, the Casimir operator $\mathbf{G}^2$ also commutes with $\tilde{\mc H}_{\text{ex-1}}$. The physical meaning of these generators is easy to see if one expresses $G^{x,y,z}$ in matrix form. For a single site, 
\begin{eqnarray}
\label{eq:hiddensu2}
G^x_i  & = & -\frac{1}{2} \left[\begin{array}{llll} 
                                & & & 1 \\
                                & & 1 & \\
                                & 1 & & \\
                                1 & & &
                             \end{array}
                         \right] = \frac{1}{2} (-\sigma^x)_{14} \oplus (-\sigma^x)_{23} \\
G^y_i & =  & \frac{1}{2} \left[\begin{array}{llll} 
                                & & & -i \\
                                & & i & \\
                                & -i & & \\
                                i & & &
                             \end{array}
                         \right] = \frac{1}{2} (\sigma^y)_{14} \oplus (-\sigma^y)_{23} \\                        
G^z_i & = &   \frac{1}{2} \left[\begin{array}{llll} 
                                -1 & &  \\
                                & 1 & & \\
                                & & -1 & \\
                                & & & 1
                             \end{array}
                         \right] = \frac{1}{2} (-\sigma^z)_{14} \oplus (\sigma^z)_{23},
\end{eqnarray}
in which the empty matrix entries are zero and we have expressed these
generators as the direct sum of two Pauli matrices, one
($\boldsymbol{\sigma}_{14}$) for the subspace of $j^z_i = \pm 3/2$
states and the other ($\boldsymbol{\sigma}_{23}$) for the subspace of
$j^z_i = \pm 1/2$ states. One intuitive way to think about these SU$(2)$
generators is that they transform the spin components in the $j^z_i =
\pm 3/2$ subspace together with $j^z_i = \pm 1/2$ subspace. This is a
global symmetry of $\tilde{\mc H}_{\text{ex-1}}$.

Now we consider the other two interactions, $\tilde{\mathcal
  H}_{\text{ex-2}}$ and $\tilde{\mathcal H}_{\text{quad}}$. We find that
the electric quadrupole-quadrupole interaction $\tilde{\mathcal
  H}_{\text{quad}}$ also commutes with $\mathbf{G}$.   On the other hand,  the ferromagnetic exchange interaction $\tilde{\mc H}_{\text{ex-2}}$ breaks this SU$(2)$ symmetry; thus
\begin{equation}
[\tilde{\mathcal H},\mathbf{G}]\propto J'.
\end{equation}
For $J'\ll J,V$, we have an approximate continuous symmetry.  

\section{Mean-field ground states}
\label{sec:sec3}

In this section, we study the ground state of the model Hamiltonian
$\tilde{\mc H}$ in Eq.~\eqref{model}.  We begin in Sec.~\ref{sec:sec31}
by getting some intuition from considering a perturbed model with strong
easy-axis or easy-plane anisotropy. This starting point also has
experimental motivation as several ordered double perovskites in
Tab.~\ref{tab:Tab1} develop such anisotropies that are driven by lattice
distortions.  Armed with the understanding of the anisotropic cases, we
proceed to analyze the  case of cubic symmetry using mean-field theory
in Sec.~\ref{sec:sec32}.  We consider briefly intermediate strength
anisotropy in Sec.~\ref{sec:interm-anis}.

In general, Curie-Weiss mean-field theory consists of decoupling all
inter-site interactions to obtain self-consistent single-site
Hamiltonians.  At zero temperature, this is equivalent to assuming a
product form for the wavefunction, i.e.
\begin{equation}
  \label{eq:20}
  |\Psi\rangle = \otimes_i |\psi_i\rangle,
\end{equation}
where the product is over sites, and $|\psi_i\rangle$ is an arbitrary
$j=3/2$ ket.  One calculates the mean-field ground state energy as the
expectation value of the Hamiltonian in this state, and minimizes it.
Thus the mean-field approximation can also be considered as a simple
variational one.

\subsection{The case with strong anisotropy}
\label{sec:sec31}

In this subsection, we add to the Hamiltonian $\tilde{\mc H}$ in
Eq. (\ref{model}) a strong anisotropic term,
\begin{equation} 
 {\mc H}_{\text{ani}} = - D \sum_i (j_i^z)^2
 \;,
\end{equation}
in which, $D$ can be positive or negative, representing easy-axis or
easy-plane anisotropy, respectively. Although this interaction is
anisotropic in spin space, it still respects the ``hidden'' global SU(2)
symmetry.

\subsubsection{Easy-axis anisotropy}
\label{sec:easy-axis-anisotropy}

Let us start with  easy-axis anisotropy $D>0$. Assuming the anisotropy
is very strong $D \gg J,J',V$ and $D \ll \lambda$, which favors $j_i^z =
\pm 3/2$ states, we can safely project the Hamiltonian $\tilde{\mc H}$
into the latter two-dimensional subspace.  We then obtain the effective
Hamiltonian
\begin{eqnarray} {\mc H}_{\text{eff-1}} & = & \sum_{\langle ij \rangle
    \in \text{XZ,YZ}} \left[\left(\frac{J}{4}+\frac{J'}{2}\right)
    \boldsymbol{T}_i \cdot \boldsymbol{T}_j - J' {T}_i^z
    {T}_j^z\right ]
  \nonumber \\
  & & + {\mc N}\left(-\frac{J}{4}+\frac{3J'}{2} + \frac{11V}{12}\right),
\label{eq:projz}
\end{eqnarray}
where we have introduced pseudospin-$1/2$ operators $\boldsymbol{T}_i$ acting on the basis $j_i^z = \pm 3/2$ with $T_i^z = \pm 1/2$ corresponding to $j_i^z =\pm 3/2$, respectively.
And ${\mc N}$ is the total number of sites. In the reduced space, the original spin vector reduces to the pseudospin in the following way:
\begin{equation}
(j^x,j^y,j^z) \Rightarrow 3(0,0,T^z)
\;.
\end{equation} 
Notice that after this projection the interaction on the horizontal
bonds in XY planes disappear in the effective Hamiltonian ${\mc
  H}_{\text{eff-1}}$. This can be understood in terms of the original
orbital picture as the easy-axis anisotropy lifts the degeneracy of
$t_{2g}$ triplets, favoring $xz$ and $yz$ orbitals to be occupied. As a
result, {\sl the above effective Hamiltonian is operating on a bond-depleted
fcc lattice, which is in fact a unfrustrated bipartite bcc lattice}.  

In
${\mc H}_{\text{eff-1}}$ because of the in-plane anisotropy introduced by
the FM exchange, the ground state of ${\mc H}_{\text{eff-1}}$ is
``antiferromagnetically'' ordered in the $(T^x,T^y)$ plane with an
ordering wavevector $\bs{Q} = 2\pi (001)$.  We denote this as the AFM
state.  The corresponding mean-field
ground state is just the direct product,
\begin{equation} 
|\Psi(\phi)\rangle=\prod_i|\psi_i(\phi)\rangle,
\label{eq:state_aniz1}
\end{equation}
where
\begin{equation} 
| \psi_i (\phi)\rangle = \frac1{\sqrt2}[| j_i^z=\tfrac{3}{2} \rangle + (-)^{2z_i}  e^{i \phi}|j_i^z=  \tfrac{3}{2}  \rangle].
\label{eq:state_aniz2}
\end{equation}
with an arbitrary phase $\phi$. The arbitrariness of the phase comes
from the U$(1)$ symmetry of the projected effective Hamiltonian
Eq.~\eqref{eq:projz}. However, as discussed in the previous section, the
continuous symmetry in the original Hamiltonian is broken completely
when $J'\neq 0$. Therefore, the U(1) symmetry of Hamiltonian in
Eq.~\eqref{eq:projz} is a by-product of the projection. Because we are
in the subspace of $j^z = \pm 3/2$, the orbital occupation is
\begin{equation}
(\langle \tilde{n}_{i,yz}  \rangle, \langle \tilde{n}_{i,xz}  \rangle, \langle\tilde{n}_{i,xy} \rangle) = (1/2,1/2,0)  .
\end{equation}
It is also important to note that the ground state in
Eq.~\eqref{eq:state_aniz2} is not a conventional N\'{e}el state as it
has a vanishing static magnetic dipole moment,
\begin{equation}
\langle \Psi | {\bs j}_i| \Psi \rangle = 0
\;!
\end{equation}
The $\phi$ dependence only shows up in the spin operators of a specific orbital,
\begin{eqnarray} 
\langle\Psi | \tilde {\bs S}_{i,yz} | \Psi \rangle &=& -\frac{1}{4}(-)^{2z_i}(\cos{\phi},\sin{\phi},0 ),\\
\langle\Psi | \tilde{\bs S}_{i,xz} | \Psi \rangle &=& \frac{1}{4}(-)^{2z_i}(\cos{\phi},\sin{\phi},0 ), \\
\langle\Psi | \tilde{\bs S}_{i,xy} | \Psi \rangle &=&  (0,0,0).
\end{eqnarray}

\subsubsection{Easy-plane anisotropy}
\label{sec:easyplane}

Now we consider easy-plane anisotropy $D<0$.  We also assume the
anisotropy is very strong $|D| \gg J,J',V$ and $|D| \ll \lambda$, which
favors $j_i^z = \pm 1/2$ states and obtain the effective Hamiltonian
after projection into the $j_i^z =\pm 1/2$ subspace,
\begin{eqnarray}
{\mc H}_{\text{eff-2}} & = & \sum_{\langle ij \rangle \in \text{XY} } \frac{4}{9}   ( J \boldsymbol{T}_i \cdot \boldsymbol{T}_j + J' T_i^z T_j^z )+
 \sum_{\langle ij \rangle \in \text{XZ} }  \left[ \frac{J}{36} \boldsymbol{T}_i \cdot \boldsymbol{T}_j \right.
 \nonumber \\
&+&\left. J'(-\frac{1}{6} T_i^x T_j^x + \frac{5}{18} T_i^y T_j^y +\frac{1}{6} T_i^z T_j^z )\right]  \nonumber \\
&+& \sum_{\langle ij \rangle \in \text{YZ} } \left[ \frac{J}{36} \boldsymbol{T}_i \cdot \boldsymbol{T}_j  + J' ( \frac{5}{18} T_i^x T_j^x -\frac{1}{6} T_i^y T_j^y  \right.
\nonumber \\
&+& \left. \frac{1}{6} T_i^zT_j^z) \right] + {\mc N} (- \frac{J}{4}  + \frac{3J'}{2}  + \frac{11V}{12} )
\;.
\end{eqnarray}
Here the pseudospin-$1/2$ operator $\boldsymbol{T} $ is acting on the subspace of $j^z = \pm 1/2$ with $T^z = \pm 1/2 $ representing $j^z = \pm 1/2$, respectively. In the reduced spin space, the original spin vector is reduced to the pseudospin in the following way:
\begin{equation}
(j^x,j^y,j^z) \Rightarrow (2 T^x, 2 T^y,T^z)\label{eq:3}
\;.
\end{equation} 

We can now find the mean-field ground state of this Hamiltonian.  For
an effective $S=1/2$ model of this type, this is equivalent to the
classical approximation.  Classically, we can find the minimum energy
states by the Luttinger-Tisza method.  This amounts to looking for the
eigenvalues of the spin Hamiltonian regarded as a quadratic form, and
seeking a classical spin solution which is built of a superposition
only of those eigenvectors which have minimum energy eigenvalues.  The
result in this case is that, for $0<J'<J$, there are two classes of
solution, all collinear spin states.  These are: (i) states with ${\bs
  Q}=2\pi(100)$ and the pseudo-spin axis in the $yz$ plane, and (ii)
states with ${\bs Q}=2\pi(010)$ and the pseudo-spin axis in the $xz$
plane.   As for the easy-axis case, there is an accidental degeneracy
of spin orientations within the plane normal to ${\bs Q}$.  Note that while
the pseudospin orients freely along a circle in this plane, the
magnetization orients along an ellipse due to the factor of 2 in
Eq.~\eqref{eq:3}.   One readily expresses the ground state in the unprojected
Hilbert space.  For example, taking ${\bs Q} = 2\pi (010)$ and
pseudospin pointing along $x$ direction, then
\begin{equation}
\label{eq:easyplanstate}
|\Psi\rangle = \prod_i | \psi_i \rangle
\end{equation}
with
\begin{equation}
\label{eq:easyplanestate}
| \psi_i \rangle = \frac{1}{\sqrt{2}}  [   | j_i^z = \tfrac{1}{2} \rangle +  (-)^{2y_i}  | j_i^z = -\tfrac{1}{2} \rangle ]
\;.
\end{equation}
This is once again an antiferromagnetic state, and to distinguish it
from the one which obtains for Ising anisotropy, we denote it AFM'.
The defining difference of the AFM' and the AFM state discussed
previously is that, the former has a non-zero dipole moment, while, at
least within mean field theory, the latter does not.  

\subsection{The cubic case}\label{sec:sec32}

Having understood the cases with strong easy-axis and easy-plane
anisotropies, let us now turn to the Hamiltonian $\tilde{\mc H}$ with
cubic lattice symmetry in Eq.~\eqref{model}. Both ground states of the
Hamiltonian with strong easy-axis or easy-plane anisotropy comprise
two-sublattice structure with an ordering wavevector equivalent to ${\bs
  Q} = 2\pi (001)$.  It is therefore natural to guess that the same two
sublattice structure is also obtained in the cubic case.  While we have not
proven this, we have investigated more general mean-field ground states,
allowing for much larger unit cells, but found in every case that the
minimum energy is found for the two-sublattice configuration.
Therefore, in what follows, we assume the two sublattice structure with
ordering ${\bs Q} = 2\pi (001)$ (which is equivalent to $2\pi (100)$ and
$2\pi(010)$ in the cubic case).  We make no further assumptions, and
minimize the energy with respect to an arbitrary wavefunction on each of
the two sublattices.  The resulting variational phase diagram is
depicted in Fig.~\ref{fig:phase_diagram}.

\subsubsection{Antiferromagnetic (AFM) state}
\label{sec:antif-afm-state}

In Fig.~\ref{fig:phase_diagram}, for small $J'/J$ and $V/J$, we find a
phase, denoted AFM, which is the natural continuation of the AFM
phases encountered in the anisotropic limits.  Here, as in those
cases, the states on the two sublattices are simply related by a time
reversal transformation, and indeed the ground state has the same form
as that found in the easy-axis case,
Eqs.~\eqref{eq:state_aniz1},\eqref{eq:state_aniz2}.  The appearance of
time-reversed pairs of sites is natural, since the largest
interaction, $\tilde{\mc H}_{\text{ex-1}}$, is dominated by the
spin-spin exchange term.  Interestingly, one finds that the ground
state has a {\sl continuous} degeneracy: the phase $\phi$ in
Eq.~\eqref{eq:state_aniz2} can be arbitrary. Since the Hamiltonian
with non-vanishing $J'$ has no continuous symmetry, this degeneracy
appears to be accidental. Since it has the same form as we found in
Sec.~\ref{sec:easy-axis-anisotropy}, we continue to use the label AFM
here for this state (and in Fig.~\ref{fig:phase_diagram}).

\subsubsection{Ferromagnetic 110 (FM110) state}
\label{sec:ferr-110-fm110}

With large $J'/J$ and $V/J$, the orbital-orbital interaction has more
weight in the Hamiltonian $\tilde{\mc H}$, and the nature of the ground
state changes.  One should note that even the pure orbital-orbital
interaction is not trivial and classical, since the orbital occupation
numbers no longer commute after projection down to the $j=3/2$
quadruplets.  However, in mean-field theory one may still treat the
expectation values classically.  Note that the largest terms in the
orbital-orbital interaction  are those which are diagonal in the
orbital basis, namely, the second term in Eq.~\eqref{exchangeFM2} and
the second term in Eq.~\eqref{quadrupole2}. To minimize the diagonal
orbital interaction like $\tilde{n}_{i,xz} \tilde{n}_{j,xz}$, a
schematic recipe is to maximize $\tilde{n}_{i,xz} $ while minimizing
$\tilde{n}_{j,xz} $.  This is necessary because one cannot minimize both
$\tilde{n}_{i,xz} $ and $\tilde{n}_{j,xz} $ simultaneously, since, due
to the single-occupancy constraint, the other diagonal terms such as
$\tilde{n}_{i,yz} \tilde{n}_{j,yz}$ would then be increased.  Since the
occupation numbers of the same orbital must be taken different on
different sites, and these occupation numbers are time-reversal
invariant, the states on the two sublattices cannot be time-reversed
counterparts.  Consequently, there is a competition between the
orbital-orbital interactions ($\tilde{\mc H}_{\text{ex-2}}$ and
$\tilde{\mc H}_{\text{quad}}$) and the nearest neighbor
antiferromagnetic exchange interaction ($\tilde{\mc
  H}_{\text{ex-1}}$).  In  the large $J'/J$ and $V/J$ region, when the
orbital interactions dominate,  we find however that time-reversal
symmetry is still broken, and since these states are not composed of
time-reversed pairs, the result is an uncompensated net ferromagnetic
moment.  

In the majority of phase space, we find the ground state is
characterized by three parameters, $r$, $\phi_1$ and $\phi_2$,
\begin{widetext}

\begin{eqnarray}
|\psi_A \rangle_{\text{FM110}} & = &  \frac{r}{
\sqrt{2}}  ( e^{i\phi_1} |j^z=\tfrac{1}{2} \rangle + e^{i(\phi_2-\phi_1)}|j^z= -\tfrac{1}{2} \rangle  ) + \sqrt{\frac{1-r^2}{2}}  \left( e^{i \phi_2} |j^z=\tfrac{3}{2} \rangle +|j^z=-\tfrac{3}{2} \rangle \right) \\
|\psi_B \rangle_{\text{FM110}} & = &     \frac{r}{
\sqrt{2}} ( - e^{-i\phi_1} |j^z=\tfrac{1}{2} \rangle + i e^{i(\phi_1 -\phi_2)}|j^z=-\tfrac{1}{2} \rangle )   +   \sqrt{\frac{1-r^2}{2}}     (-i e^{-i \phi_2 } |j^z=\tfrac{3}{2}  \rangle +|j^z= -\tfrac{3}{2} \rangle )
\;,
\label{eq:stateFM1}
\end{eqnarray}
in which, ``A'' and ``B'' represent the two sublattices, and $r, \phi_1$ and $\phi_2$ are determined by minimizing the mean field energy. Note that in
Eq.~\eqref{eq:stateFM1} the three parameters $r, \phi_1$ and $\phi_2$
are uniquely determined by $J'/J$ and $V/J$. So the orbital occupations
can be readily generated,
\begin{eqnarray}
(\langle \tilde{n}_{A,yz}  \rangle, \langle \tilde{n}_{A,xz}  \rangle, \langle\tilde{n}_{A,xy} \rangle )_{\text{FM110}} &=&  (\frac{1}{2} -\frac{r^2}{3}-\frac{r\sqrt{1-r^2} }{\sqrt{3}} \cos{\phi_1},\frac{1}{2} -\frac{r^2}{3}+\frac{r \sqrt{1-r^2}}{\sqrt{3}} \cos{\phi_1}, \frac{2r^2}{3}), \\
(\langle \tilde{n}_{B,yz}  \rangle, \langle \tilde{n}_{B,xz}  \rangle, \langle\tilde{n}_{B,xy} \rangle )_{\text{FM110}} &=&  (\frac{1}{2} -\frac{r^2}{3}+\frac{r\sqrt{1-r^2} }{\sqrt{3}} \cos{\phi_1},\frac{1}{2} -\frac{r^2}{3}-\frac{r\sqrt{1-r^2} }{\sqrt{3}} \cos{\phi_1}, \frac{2r^2}{3}) .
\end{eqnarray}

It is interesting to see the spin vectors of two sublattices are symmetric about $[1\bar{1}0]$ direction,
\begin{eqnarray}
\langle {\bs j}_A \rangle_{\text{FM110}} &=& r \left(   \sqrt{3-3r^2} \cos{(\phi_1-\phi_2)} + r \cos{(2\phi_1-2\phi_2)},    \sqrt{3-3r^2} \sin{(\phi_1-\phi_2)} -r  \sin{(2\phi_1-2\phi_2)}, 0 \right) \\
\langle {\bs j}_B \rangle_{\text{FM110}} &=& r \left(- \sqrt{3-3r^2} \sin{(\phi_1-\phi_2)} + r  \sin{(2\phi_1-2\phi_2)},  -\sqrt{3-3r^2} \cos{(\phi_1-\phi_2)} - r \cos{(2\phi_1-2\phi_2)},0    \right)
\end{eqnarray}
so the system has a non-vanishing net spin polarization, that is 
\begin{equation}
    \tfrac{1}{2} \langle {\bs j}_A + {\bs j}_B\rangle_{\text{FM110}}  =    \frac{r}{2} \left[ \sqrt{3-3r^2} (\cos{(\phi_1-\phi_2)} -  \sin{(\phi_1-\phi_2)} ) + r  (\cos{(2\phi_1-\phi_2)} + \sin{(2\phi_1-\phi_2)} )\right] (1,-1,0)\;.
\end{equation}
This direction of polarization is equivalent to $[110]$ by a 90 degree
rotation, so we denote this a FM110 state.  It occupies the
corresponding region in Fig.~\ref{fig:phase_diagram}.

\subsubsection{Ferromagnetic 100 (FM100) state}
\label{sec:ferr-100-state}

Between the AFM and FM110 states,  a narrow region
of intermediate phase intervenes (see
Fig.~\ref{fig:phase_diagram}). Numerically we find this phase is
characterized by two parameters $r_1$ and $r_2$,
\begin{eqnarray}
|\psi_A \rangle_{\text{FM100}} & = &  \frac{r_1}{
\sqrt{2}} (  |j^z=\tfrac{1}{2} \rangle + |j^z= -\tfrac{1}{2} \rangle ) + \sqrt{\frac{1-r_1^2}{2}}  \left(  |j^z=\tfrac{3}{2} \rangle +|j^z=-\tfrac{3}{2} \rangle \right) \\
|\psi_B \rangle_{\text{FM100}} & = &   \frac{r_2}{
\sqrt{2}} (-  |j^z=\tfrac{1}{2} \rangle + |j^z=-\tfrac{1}{2} \rangle  ) + \sqrt{\frac{1-r_2^2}{2}} (-|j^z=\tfrac{3}{2} \rangle +|j^z= -\tfrac{3}{2} \rangle )
\;.
\label{eq:FM100}
\end{eqnarray}
The parameters $r_1$ and $r_2$ are determined by $J'/J$ and $V/J$ and
in this intermediate phase $r_1 \neq r_2$. A second ground state is
obtained then by interchanging $r_1$ and $r_2$. The orbital
occupation numbers and spin vectors are give by
\begin{eqnarray}
(\langle \tilde{n}_{A,yz}  \rangle, \langle \tilde{n}_{A,xz}  \rangle, \langle\tilde{n}_{A,xy} \rangle )_{\text{FM100}} &=&  (\frac{1}{2} -\frac{r_1^2}{3}-\frac{r_1}{\sqrt{3}} \sqrt{1-r_1^2},\frac{1}{2} -\frac{r_1^2}{3}+\frac{r_1}{\sqrt{3}} \sqrt{1-r_1^2}, \frac{2r_1^2}{3}), \\
(\langle \tilde{n}_{B,yz}  \rangle, \langle \tilde{n}_{B,xz}  \rangle, \langle\tilde{n}_{B,xy} \rangle )_{\text{FM100}} &=&  (\frac{1}{2} -\frac{r_2^2}{3}+\frac{r_2}{\sqrt{3}} \sqrt{1-r_2^2} ,\frac{1}{2} -\frac{r_2^2}{3}-\frac{r_2}{\sqrt{3}} \sqrt{1-r_2^2} , \frac{2r_2^2}{3}) 
\end{eqnarray}
and
\begin{eqnarray}
\langle {\bs j}_A \rangle_{\text{FM100}} &=&  \left(r_1^2 +  r_1 \sqrt{3-3r_1^2} , 0, 0 \right) \\
\langle {\bs j}_B \rangle_{\text{FM100}} &=&  \left(r_2^2  -  r_2 \sqrt{3-3r_2^2}, 0, 0    \right)\;.
\end{eqnarray}
We see that the net spin polarization is along the $[100]$ direction.
 Due to cubic symmetry, all possible $[ 100]$ directions
 are possible.  By analogy with the previous phase, we denote this
 phase FM100.  It occupies the narrow region shown in Fig.~\ref{fig:phase_diagram}.

\end{widetext}

\subsubsection{Transitions}
\label{sec:transitions}

\begin{figure}[htp]
\subfigure{\includegraphics[width=6.5cm]{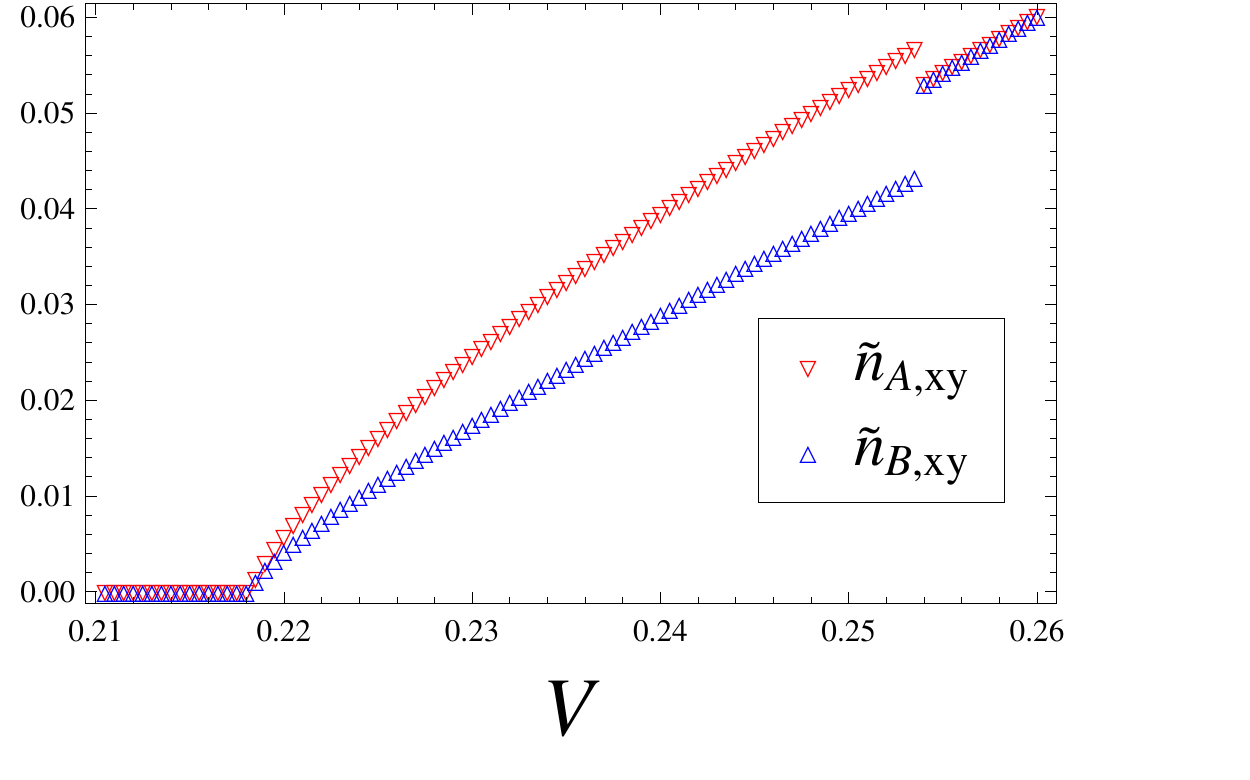}}
\subfigure{\includegraphics[width=6.5cm]{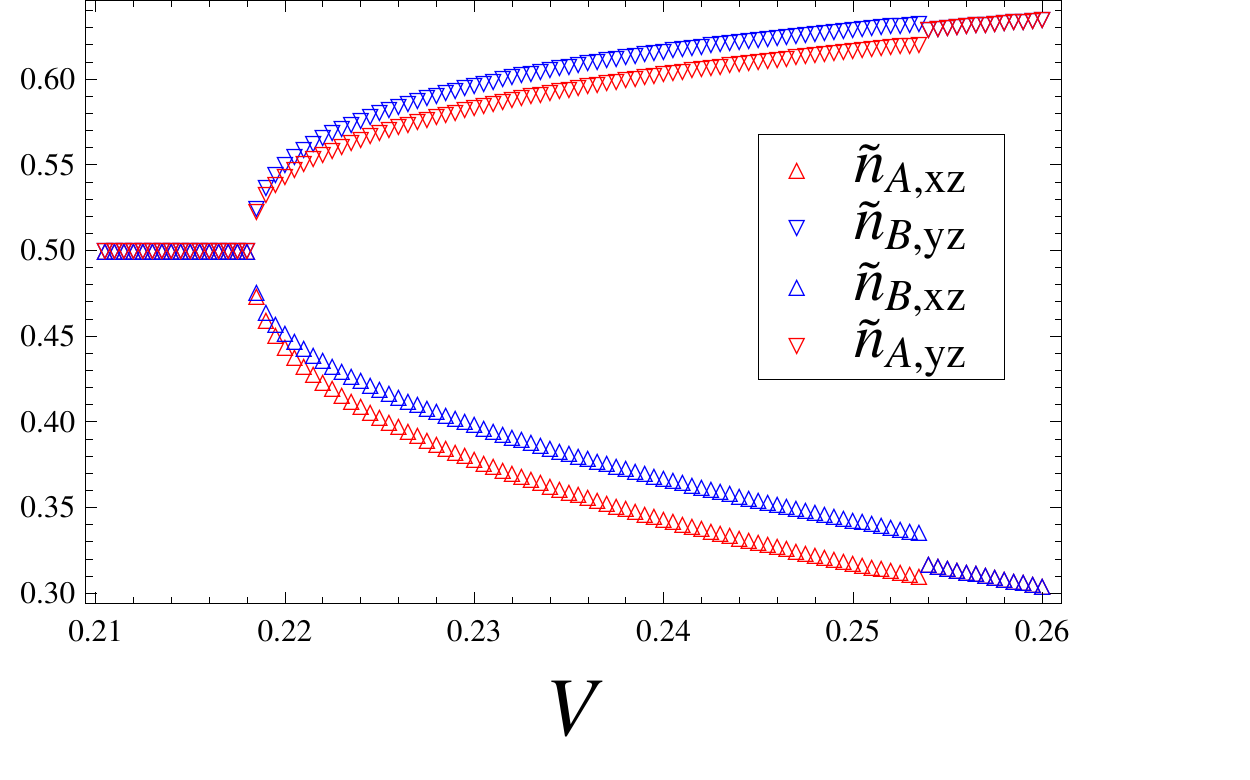}}
\subfigure{\includegraphics[width=6.5cm]{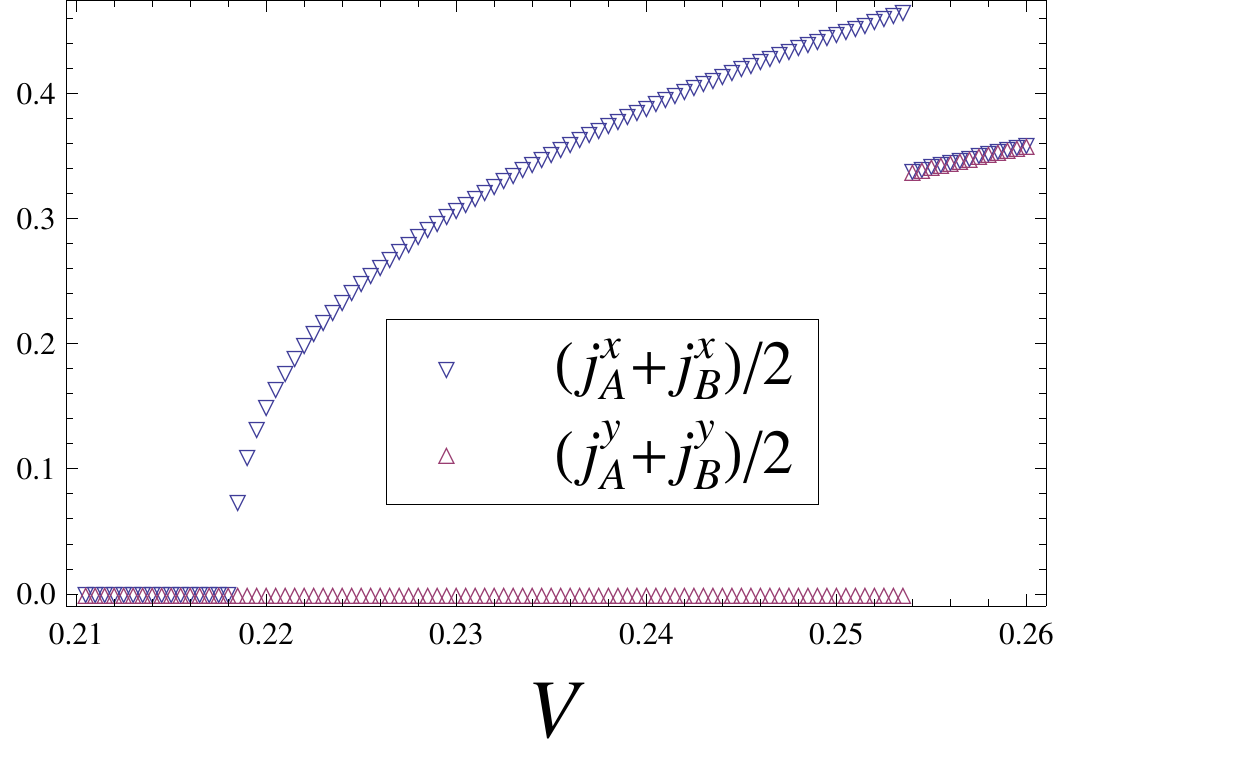}}
\caption{ (Color online) Upper graph: $\tilde{n}_{xy}$ versus $V$;
  middle graph: $\tilde{n}_{xz}$ and $\tilde{n}_{yz}$ versus $V$;
  lower graph: the net spin polarization per site versus $V$. In all
  three graphs, $J=1$ and $J'=0.2$, and $T=0$. The assignment of A and
  B sublattices is arbitrary. Here we take the choice given in the
  text. }
\label{fig:orbitalspin}
\end{figure}

The intermediate FM100 state in Eq.~\eqref{eq:FM100} is smoothly
connected to the AFM state, which is obtained by setting $r_1=r_2=0$.
By contrast, it cannot be connected to the FM110 state. This indicates
that the transition between FM100 to AFM is continuous while the
transition from FM100 to FM110 is first-order.  Indeed, this can also
be clearly seen from the behavior of the spin and orbital order
parameters across these transitions (see
Fig.~\ref{fig:orbitalspin}). Both the spin and orbital order
parameters jump when the system goes from FM100 phase to FM110 phase.

The approach to the line $J'=0$, while not a transition {\sl per se},
does represent a change of behavior.  In particular, on this line, the
hidden SU(2) symmetry is restored, and new ground states may be
obtained from the above three phases by SU(2) rotations.  This allows,
for instance, for AFM states to develop with non-vanishing magnetic
dipole order in the ground state.  However, for arbitrarily small $J'$
the SU(2)-induced degeneracy is broken and the results quoted above
hold. 

\subsection{Intermediate anisotropy}
\label{sec:interm-anis}

We briefly address here the evolution of the ground states with
increasing $|D|$ between the cubic and strongly anisotropic limits.
For easy-axis anisotropy, $D>0$, this process is relatively simple.
The AFM phase (with ${\bs Q}=2\pi(001)$) is favored by this sign of anisotropy, and therefore,
with increasing $D$, it expands at the expense of the FM110 and FM100
states.  Indeed, for very large $D$, ferromagnetic states occur only
for unphysically large $J'$ and $V$.  

In the case of easy-plane anisotropy, $D<0$, the phase diagram is more
subtle.  For weak $|D|$, the main effect is to break the symmetry
between the formerly equivalent $[ 100]$ wavevectors.
In this case, states with minimal $\langle j_z^2 \rangle$ are favored,
which prefers ${\bs Q}=2\pi(100), 2\pi(010)$ rather than ${\bs Q}=2\pi(001)$.  

If we begin in the AFM state for $D=0$, this aligns the
pseudospin in the plane normal to this wavevector.  The phase
degeneracy which obtains for the cubic case is broken by the
anisotropy, and a definite alignment is obtained.  Moreover, as states
with $j_z=\pm 1/2$ are increasingly mixed into the ground state, a
non-vanishing dipole moment, proportional to the pseudospin, is
induced.  The magnitude of this staggered magnetization grows
continuously with $|D|$, eventually as $D\rightarrow -\infty$,
approaching the value obtained in Sec.~\ref{sec:easyplane}.  This
local moment is oriented in the plane normal to ${\bs Q}$, and can take values
distributed over an {\sl ellipse} in this plane.  In the large $|D|$
limit, this ratio of the major (perpendicular to ${\bs Q}$ and to $z$) and minor
($z$) axes of the ellipse approaches 2, corresponding to the accidental
degeneracy discussed in Sec.~\ref{sec:easyplane} .  Because the state
for non-zero $D$ evolves smoothly into this limit, and has a non-zero
local moment, we denote it an AFM' state, following the earlier
notation.

\begin{figure}[htp]
\includegraphics[width=6.5cm]{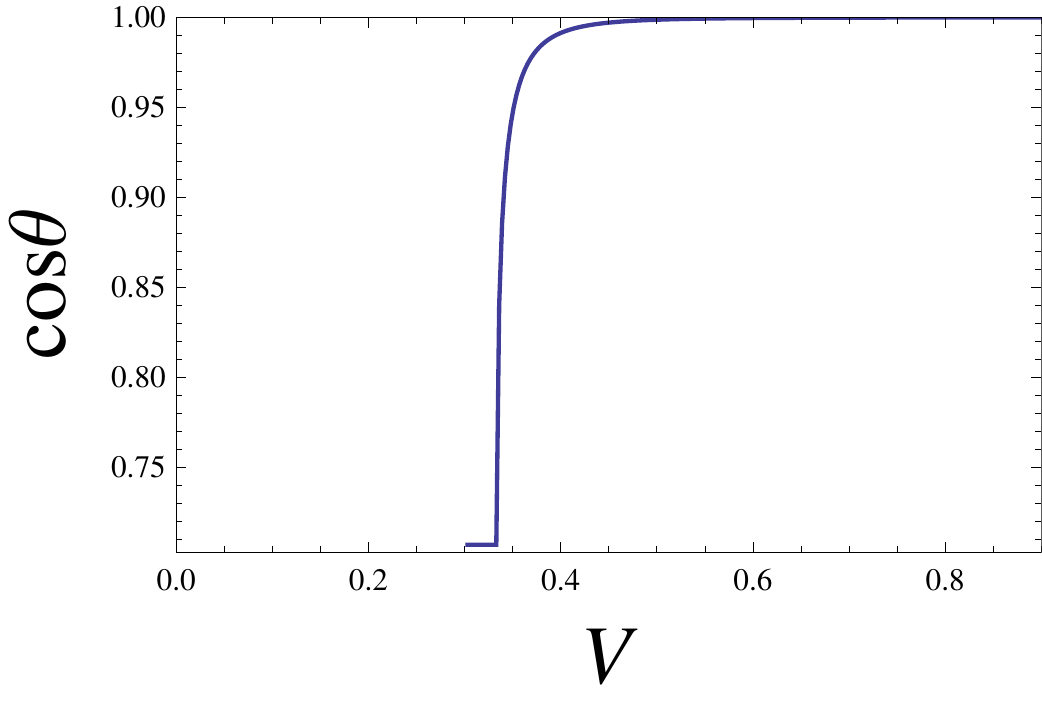}
\caption{ (Color online) The orientation of the net spin polarization
  for the FM110$^*$ phase, at $T=0$. $\theta$ is angle between the net
  spin polarization and nearest $[110]$ direction.  In the figure,
  $J'=0.2, D=-0.05, J=1$. $\cos {\theta}$ increases from $1/\sqrt{2}$
  for the FM100 phase to $1$ for the FM110$^*$ phase as $V$ goes
  through the phase transition point.}
\label{fig:orientation}
\end{figure}

\begin{figure}[htp]
\includegraphics[width=6cm]{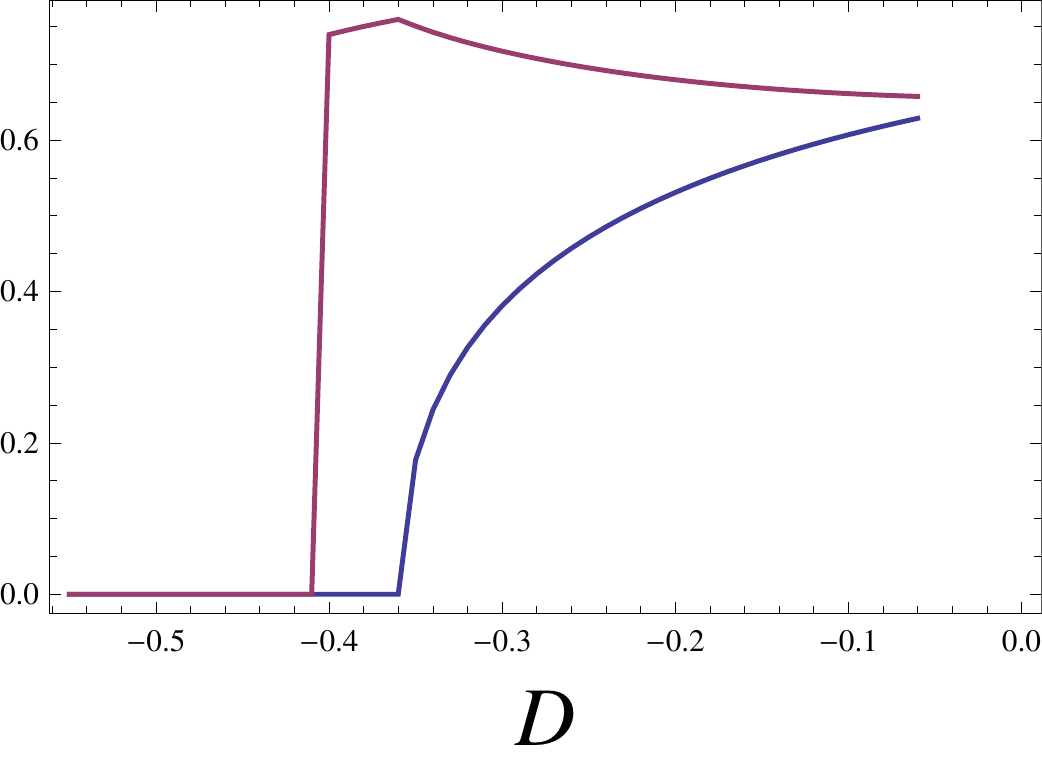}
\caption{ (Color online) The net polarization versus the easy plane
  anisotropy, at $T=0$.  The upper curve (in red) is for the $y$
  component of the net polarization. And the lower curve (in blue) is
  for the $z$ component of the net polarization. The ordering
  wavevector is ${\bs Q} = 2\pi (100)$.  In the graph, $J'=V=0.4$ and
  $J=1$.  When $0<|D|\lesssim 0.36$, the system is in FM110$^*$ phase;
  when $0.36<|D|<0.41$, the system is in the FM100 phase; when $|D|\gtrsim
  0.41$, the system is in AFM' phase.}
\label{fig:orientation1}
\end{figure}

Beginning in the FM110 state at $D=0$, one observes two subsequent
transitions.  First, small $|D|$ orients the magnetization normal to ${\bs Q}$.
For concreteness consider ${\bs Q}=2\pi(100)$, in which case one obtains a
ferromagnetic magnetization of the form ${\bs m}=(0,m_1,m_2)$. For
$D=0^-$, $m_1=m_2$, but subsequently $m_2$ decreases such that
$m_2<m_1$.  We denote this state FM110*.
An example for the orientation of net polarization in the FM110* by varying $V$ is given
in Fig.~\ref{fig:orientation}.
 Eventually once some critical
anisotropy is reached, $m_2$ vanishes continuously.  At this point the
magnetization is aligned along the $(010)$ axis.  For yet larger
anisotropy, eventually the ferromagnetic magnetization vanishes
entirely, and the ground state switches to the AFM' state. An example of the 
transitions from FM110* to FM100 then to AFM' by varying the easy-plane anisotropy $D$
is given in Fig.~\ref{fig:orientation1}.

Finally, starting in the FM100 state, the magnetization immediately
switches to the $(010)$ direction.  This is the same phase as the
intermediate phase observed starting from the FM110 phase.  Thus with
further increase in anisotropy, the ground state switches to the AFM'
state.  

One may also visualize the evolution of the ground states with
anisotropy by considering planar phase diagrams at fixed $D$.  With
increasing positive $D$ (Ising anisotropy), the AFM state is stabilized,
and simply expands in the $J'-V$ plane, pushing the FM100 and FM110
states outward.  For increasing negative $D$, apart from the fact that
the AFM and FM110 states evolve into the AFM' and FM110* states, the
behavior is similar: the AFM' state expands at the expense of the
ferromagnetic states.

\section{Multipolar orders and $T>0$ behavior}
\label{sec:sec4}

\subsection{Order parameters}
\label{sec:order-parameters}

In this section, we extend the analysis of the previous section to
non-zero temperature.  To do so, we employ the usual extension of mean
field theory to include thermal fluctuations.  To characterize the
phases encountered in this treatment, it is natural to introduce several
types of order parameter.  First, on a single site $i$, we may measure
the dipole moment, which is proportional to ${\bs j}_i$.  However, we
may also measure the next two multipoles: the quadrupole moment,
proportional to
\begin{equation}
  \label{eq:4}
  Q_{i}^{\mu\nu} = \left\langle j_i^\mu j_i^\nu\right\rangle -
  \frac{j(j+1)}{3} \delta^{\mu\nu}, 
\end{equation}
and the octupole moment
\begin{equation}
  \label{eq:5}
  O_i^{\mu\nu\lambda} = \left\langle j_i^\mu j_i^\nu j_i^\lambda
  \right\rangle.  
\end{equation}
A typical magnetic state has a non-vanishing local dipole moment,
which inevitably induces some higher multipole order parameters (see
below).  However, one sometimes encounters purely multipole states, in
which $\langle j_i^\mu\rangle=0$ but $Q_i^{\mu\nu}$ and/or
$O_i^{\mu\nu\lambda}$ are/is non-vanishing.  

\begin{table}[ht]
\centering 
\begin{tabular}{|l|c|c|} 
\hline
Moment  & Symmetry & Operator  \\ 
\hline \hline 
Dipole  & $\Gamma_4$ & $M^x = j^x$  \\  
        &            & $M^y = j^y$ \\
        &            & $M^z = j^z$ \\
        \hline
Quadrupole & $\Gamma_3$ & $Q^{3z^2} = [3(j^z)^2- {\bm j}^2]/\sqrt{3} $ \\
&&                         $Q^{x^2-y^2} = ( j^x)^2-(j^y)^2  $  \\ 
&             $\Gamma_5$ & $Q^{xy} = \overline{j^x j^y}/2  $ \\
&&                         $Q^{yz} = \overline{j^y j^z}/2  $ \\
&&                         $Q^{xz} = \overline{j^z j^x}/2  $   \\
\hline                              
Octupole & $\Gamma_2$ &   $ T_{xyz} = \sqrt{15}/6 \overline{j^x j^y j^z}  $ \\
          & $\Gamma_4$ &   $ T^x_{\alpha} = (j^x)^3 - [\overline{j^x (j^y)^2} + \overline{(j^z)^2 j^x}]/2 $ \\
          &&               $ T^y_{\alpha} = (j^y)^3 - [\overline{j^y (j^z)^2} + \overline{(j^x)^2 j^y}]/2 $ \\
          &&               $ T^z_{\alpha} = (j^z)^3 - [\overline{j^z( j^x)^2} + \overline{(j^y)^2 j^z}]/2 $ \\
          & $\Gamma_5$ &   $ T^x_{\beta} = \sqrt{15}[\overline{j^x (j^y)^2} - \overline{(j^z)^2 j^x}]/6 $ \\
          &&               $ T^y_{\beta} = \sqrt{15}[\overline{j^y (j^z)^2} - \overline{(j^x)^2 j^y}]/6 $ \\
          &&               $ T^z_{\beta} = \sqrt{15}[\overline{j^z (j^x)^2} - \overline{(j^y)^2 j^z}]/6 $ \\
\hline
\end{tabular}
\caption{Multipole moments within a cubic $\Gamma_8$ quartet. Bars over symbols indicate
the sum with respect to all the possible permutations of the indices, e.g.
$\overline{j^x (j^y)^2} = j^x (j^y)^2 + j^y j^x j^y + (j^y)^2 j^x$. Adapted from Ref.~\onlinecite{RevModPhys.81.807} and Ref.~\onlinecite{JPSJ.67.941}.}
\label{tab:Tab2}
\end{table}

The components of these tensor can be decomposed into irreducible
representations of the cubic group (characterizing the symmetry of the
ideal double perovskite structure).  This decomposition is described
fully in Table~\ref{tab:Tab2}.  Here we note in particular the
two-dimensional $\Gamma_3$ representation
\begin{eqnarray}
  \label{eq:6}
  Q_{i}^{3z^2} &  = & \frac{1}{\sqrt{3}}\left\langle 3(j_i^z)^2 - j(j+1)\right\rangle , \nonumber \\
  Q_i^{x^2-y^2} & = & \left\langle (j_i^x)^2 - (j_i^y)^2\right\rangle,
\end{eqnarray}
which are analogous to the $e_g$ orbitals in atomic physics.  The
remaining three independent components of $Q_i^{\mu\nu}$ ($j_i^x j_i^y
+ j_i^y j_i^x$ etc.) form a three-dimensional representation analogous
to the $t_{2g}$ orbitals, but do not appear in our analysis.

Another important way to break up the tensor order parameters is into
combinations which appear in the spin Hamiltonian.  Specifically,
these are the orbital occupation operators,
$\tilde{n}_{i,yz},\tilde{n}_{i,xz},\tilde{n}_{i,xy}$, and the
orbitally-resolved spin operators,
$\tilde{S}^\mu_{i,yz},\tilde{S}^\mu_{i,xz},\tilde{S}^\mu_{i,xy}$.
These can be expressed in terms of the multipoles describe above.  For
the occupation numbers,
\begin{eqnarray}
  \label{eq:7}
  \tilde{n}_{i,yz} & = & \frac{1}{3} + \frac{1}{6\sqrt{3}}Q_i^{3z^2} -
 \frac{1}{6} Q_i^{x^2-y^2}, \nonumber \\
  \tilde{n}_{i,xz} & = & \frac{1}{3} + \frac{1}{6\sqrt{3}}Q_i^{3z^2} +
 \frac{1}{6} Q_i^{x^2-y^2}, \nonumber \\
  \tilde{n}_{i,xy} & = & \frac{1}{3} - \frac{1}{3\sqrt{3}}Q_i^{3z^2} .
\end{eqnarray}
The orbitally-resolved spins decompose as
\begin{eqnarray}
  \label{eq:8}
\tilde{S}_{i,yz}^x & = & \frac{1}{15} j^x_i - \frac{2}{15}
T_{i,x}^\alpha \nonumber \\
\tilde{S}_{i,yz}^y & = & \frac{2}{15} j^y_i + \frac{1}{15}
  T_{i,\alpha}^y + \frac{1}{3\sqrt{15}} T_{i,\beta}^y, \nonumber \\
\tilde{S}_{i,yz}^z & = & \frac{2}{15} j^z_i + \frac{1}{15}
  T_{i,\alpha}^z - \frac{1}{3\sqrt{15}} T_{i,\beta}^z, \nonumber \\
\tilde{S}_{i,xz}^x & = & \frac{2}{15} j^x_i + \frac{1}{15}
  T_{i,\alpha}^x - \frac{1}{3\sqrt{15}} T_{i,\beta}^x, \nonumber \\
\tilde{S}_{i,xz}^y & = & \frac{1}{15} j^y_i - \frac{2}{15}
T_{i,\alpha}^y \nonumber \\
\tilde{S}_{i,xz}^z & = & \frac{2}{15} j^z_i + \frac{1}{15}
  T_{i,\alpha}^z + \frac{1}{3\sqrt{15}} T_{i,\beta}^z, \nonumber \\
  \tilde{S}_{i,xy}^x & = & \frac{2}{15} j^x_i + \frac{1}{15}
  T_{i,\alpha}^x + \frac{1}{3\sqrt{15}} T_{i,\beta}^x, \nonumber \\
\tilde{S}_{i,xy}^y & = & \frac{2}{15} j^y_i + \frac{1}{15}
  T_{i,\alpha}^y - \frac{1}{3\sqrt{15}} T_{i,\beta}^y, \nonumber \\
\tilde{S}_{i,xy}^z & = & \frac{1}{15} j^z_i - \frac{2}{15}
T_{i,\alpha}^z 
\end{eqnarray}

\subsection{Cubic system: phases}
\label{sec:cubic-system:-phases}

We first discuss the phases occurring in the cubic system at $T>0$. The
ground states discussed earlier are all stable to small thermal
fluctuations, and hence persist at low temperature.  Thus we expect,
broadly speaking, an antiferromagnetic (AFM) and ferromagnetic
(FM110/FM100) region at low temperature.  Of course, at temperatures
much larger than $J$, one has a disordered paramagnetic phase.
Interestingly, an additional phase appears at intermediate
temperature.  This is a non-magnetic {\sl quadrupolar ordered} phase.  

To see how this arises, we describe the mean-field procedure and its
results.  Mean field theory is formulated in the usual way.  We
self-consistently decouple interactions between different sites $i$
and $j$ as follows:
\begin{eqnarray} 
\hat{\mc O}_i \cdot \hat{\mc O}_j & \Rightarrow & \hat{\mc O}_i \cdot
\langle \hat{\mc O}_j \rangle + \langle \hat{\mc O}_i \rangle\cdot \hat{\mc
  O}_j \\ & & - \langle\hat{\mc O}_i \rangle \cdot \langle \hat{\mc O}_j \rangle.\nonumber
\;,
\end{eqnarray} 
where $\hat{\mc O}_i$ and $\hat{\mc O}_j$ are two operators at site
$i$ and $j$, respectively.  These operators are nothing but the
orbital occupation numbers and orbitally resolved spins, which are
related to the multipolar operators by Eqs.~\eqref{eq:7}-~\eqref{eq:8}.
Decoupling {\sl all} pairwise interactions between sites in this way,
we then obtain a set of single-site problems for each $j=3/2$.  Note
that these single-site problems involve not just the usual Weiss
exchange field, but also ``multipolar fields'', which act as effective
second and third order spin anisotropies.  The mean-field equations
determine self-consistent values of the orbital occupation numbers and
orbitally resolved fields.  As it is straightforward to formulate the
mean-field equations, and solve them numerically, we do not give the
details of these calculations here.

A distinct class of solutions describes each phase.  For the
antiferromagnetic phase, we find the following operators are non-zero:
\begin{eqnarray}
  \label{eq:9}
  \langle {\bm j}_i \rangle & = & \pm n (u_1,u_2,0), \\
  \langle Q_i^{3z^2}\rangle & = & q, \\
  \langle {\bm T}_{i,\alpha} \rangle  &= & \pm t_\alpha (u_1,u_2,0),
  \\
  \langle {\bm T}_{i,\beta} \rangle  &= & \pm t_\beta (-u_1,u_2,0),
\end{eqnarray}
where we have taken ${\bs Q}=2\pi(0,0,1)$ for concreteness, and the upper
and lower signs refer to the A and B sublattices, respectively.  The
parameters $n,q,t_\alpha,$ and $t_\beta$ are positive at all $T>0$ in
the AFM phase.  However, note that $n$ vanishes in the limit
$T\rightarrow 0$, in agreement with the vanishing dipole moment
discussed earlier for the AFM ground state.

In the FM110 state, the non-zero expectation values are:
\begin{eqnarray}
  \label{eq:10}
  \langle {\bm j}_i \rangle & = & m(1,1,0) \pm n (1,-1,0), \\
  \langle Q_i^{3z^2} \rangle & = & q, \\
  \langle Q_i^{x^2-y^2}\rangle & = & \mp q', \\
  \langle {\bm T}_{i,\alpha}\rangle & = & t_\alpha(1,1,0) \pm
 \tilde{t}_\alpha(1,-1,0), \\
  \langle {\bm T}_{i,\beta}\rangle & = & t_\beta(1,-1,0) \pm
 \tilde{t}_\beta(1,1,0), 
\end{eqnarray}
where again we took ${\bs Q}=2\pi(0,0,1)$ and the upper/lower signs
refer to the A/B sublattices.  In this case the parameters
$m,n,q,q',t_\alpha,\tilde{t}_\alpha,t_\beta,\tilde{t}_\beta$
are all non-zero at temperatures within the FM110 phase including
$T=0$.  

The third ordered phase dominating the phase diagram is the
quadrupolar one (For the purposes of this section, we ignore the FM100 phase, which
extends into a narrow region of ferromagnetic state with variable
polarization direction at $T>0$, as it occupies a very small volume of
the phase diagram).  In the quadrupolar state, there is only a single
non-vanishing order parameter:
\begin{equation}
  \label{eq:11}
  \langle Q_i^{x^2-y^2}\rangle  =  \mp q' .
\end{equation}

Let us discuss the symmetries of these three states.  In the AFM and
FM110 phases, time reversal symmetry is broken.  However, the net
magnetization vanishes in the AFM state.  In the AFM state, this is
guaranteed by invariance under the combined operations of translation
(such as by $(0,1/2,1/2)$, which interchanges the A and B sublattices)
and time-reversal.  No such symmetry can be combined with
time-reversal in the FM110 case.  Various point group symmetries are
also present in the AFM and FM110 phases, but we do not describe this
in detail.

In the quadrupolar case, time-reversal symmetry is unbroken, which is
sufficient to require the dipolar and octupolar order parameters to
vanish.  Only point group symmetries are broken by the quadrupolar
order.  Four-fold ($C_4$) rotations about the $x$ or $y$ axes, and
three-fold ($C_3$) rotations about $[ 111]$ axes are
broken in this state.  While the $C_4$ rotation about the $z$ axis is
also broken, the combination of this $C_4$ rotation and a translation
exchanging the A and B sublattices remains a symmetry of the
quadrupolar state.  

A standard classification scheme for quadrupolar states is to
examine the examine the eigenvalues of the $Q_i^{\mu\nu}$ matrix.
These must sum to zero because the matrix is traceless.  States in
which there are only two distinct eigenvalues, i.e. ${\rm eigs}(Q) =
\{ q,q,-2q\}$ are called {\sl nematics}, and correspond to the
situation in which one principal axis is distinguished from the other
two, which remain identical.  In the most general case, there are
three distinct eigenvalues, i.e. ${\rm eigs}(Q) = \{
q_1,q_2,-q_1-q_2\}$, with $q_1\neq q_2$.  This is called a {\sl
  biaxial nematic}, and is a state in which all three principal axes
are distinct.  The quadrupolar state obtained here is such a biaxial
nematic.  Physically, the local susceptibility in this state takes
distinct values $\chi_{\rm local}^{xx}, \chi_{\rm
  local}^{yy},\chi_{\rm local}^{zz}$ for fields along each of the
axes.  However, note from Eq.~\eqref{eq:11} that the quadrupolar order
parameter changes sign between the two sublattices.  Thus we should
properly call this state an {\sl antiferro-biaxial nematic}.  Due to
the staggered ordering, the bulk susceptibility does not distinguish
all three axes.  Instead, there are only two distinct components,
$\chi^{xx}=\chi^{yy} \neq \chi^{zz}$.  The difference between the two
components of the susceptibility serves as a simple macroscopic
means to observe quadrupolar ordering.

\subsection{Cubic system: phase diagram and transitions}
\label{sec:cubic-system:-phase}

\begin{figure}[htp]
\includegraphics[width=8cm]{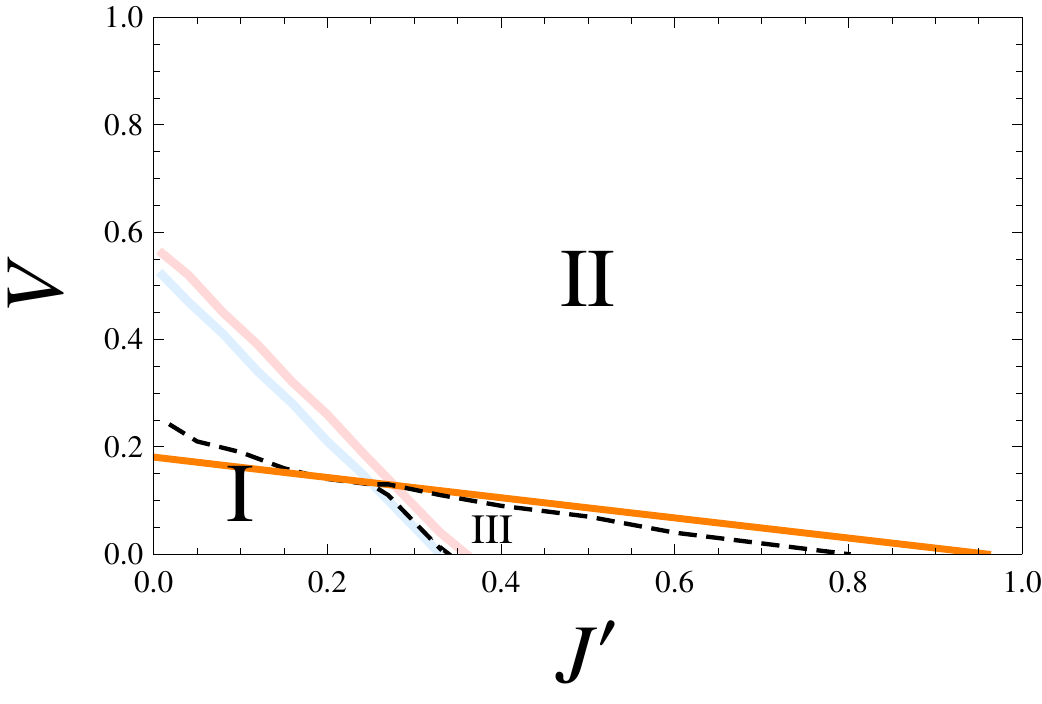}
\caption{(Color online) Zero temperature two-dimensional phase diagram
  in the cubic case (same as Fig.~\ref{fig:phase_diagram}), overlaid
  with the regions of different $T>0$ behavior. Dashed lines are obtained from 
mean field numerics. The solid line (in orange) is defined by Eq.~\eqref{eq:14}.
 In region I, there is
  a single transition to the AFM state.  In region II, the system
  supports an intermediate temperature quadrupolar ordered phase.  The
  transition from the normal state to the quadrupolar phase is second
  order.  In this region, there is a first order transition to the AFM
  phase on further cooling, for parameters such that the latter is the
  ground state.  Otherwise, the lower temperature transition is to a
  ferromagnetic (predominantly FM110) state. In region III, the system
first turns from normal phase to the AFM state then to a ferromagnetic (predominantly FM110) state on further cooling.}
\label{fig:thermal_phase}
\end{figure}

By solving the mean-field equations numerically, we have determined
the phase diagram for the cubic case.  Parts of it can be understood
analytically.  Suppose that the transitions from the high temperature
normal phase to the quadrupolar and AFM phases are second order.  This
appears to be always true for the quadrupolar phase, while it true for
the AFM for most parameters, but weakly violated in some regions. With
this assumption, we can determine the critical temperatures for these
transitions by the usual condition of marginal stability (vanishing of
the quadratic term in the Landau theory) of the free energy.  We find
that the critical temperature for the quadrupolar state is
\begin{equation}
  \label{eq:12}
  T_c^{\rm quad} = \frac{43V+18J'- 3J}{18},
\end{equation}
and that for the AFM state is
\begin{equation}
  \label{eq:13}
  T_c^{\rm AFM} = \frac{J + 10J' + \sqrt{73J^2 + 164 J J' + 100 (J')^2}}{36}.
\end{equation}
Without the assumption that the transitions are continuous, the
critical temperature could be higher.  Thus
Eqs.~(\ref{eq:12},\ref{eq:13}) give lower bounds for the transition
temperatures, strictly speaking.  Extending the two-dimensional $T=0$
phase diagram in Fig.~\ref{fig:phase_diagram} into a third dimension
of temperature, the quadrupolar phase occurs ``above'' the portion for
which $T_c^{\rm quad}> T_c^{\rm AFM}$, which implies $V>V_c$, with
\begin{equation}
  \label{eq:14}
  V_c = \frac{7J - 26J' + \sqrt{73J^2 + 164 J J' + 100 (J')^2}}{86}.
\end{equation}
The curve $V_c(J')$ defines an almost straight line in the 2d phase
diagram, as shown in Fig.~\ref{fig:thermal_phase}.  In fact, Eq.~\eqref{eq:14} slightly
underestimates $V_c$, as it assumes the normal to AFM transition is
continuous, when it is in fact weakly first order in this vicinity.
However, the true $V_c$ found numerically is only a few percent
larger.  For $V<V_c$, no quadrupolar phase occurs.  Instead, the first
ordering transition from high temperature is into the AFM state.  This
is true even when the ground state is ferromagnetic, so that in this
case (when $V<V_c$) the system first orders into the AFM state, and
then at lower temperature switches to the FM110 phase.

We now discuss the nature of the transitions.  The quadrupolar
ordering transition is, as already mentioned, continuous (see Figs.~\ref{fig:ops1030},~\ref{fig:ops3030}).  It is
described by a single scalar order parameter, (equal to $q'$ in
Eq.~\eqref{eq:11}), for each of the three $[100]$ wavevectors,
describing the associated staggered quadrupole moment.  Formally,
\begin{equation}
  \label{eq:15}
  \phi_a = (-1)^{2x_{i}^a} \langle Q^{x^2-y^2}_i \rangle,
\end{equation}
where $x_i^1=x_i, x_i^2=y_i,x_i^3=z_i$.  According to symmetry, the
Landau free energy for $\phi_a$ has the same form as that for an O(3)
magnetic transition with cubic anisotropy.  Beyond mean field theory, this is
believed to support a three dimensional transition in the O(3)
universality class.  

\begin{figure}[htp]
\includegraphics[width=8cm]{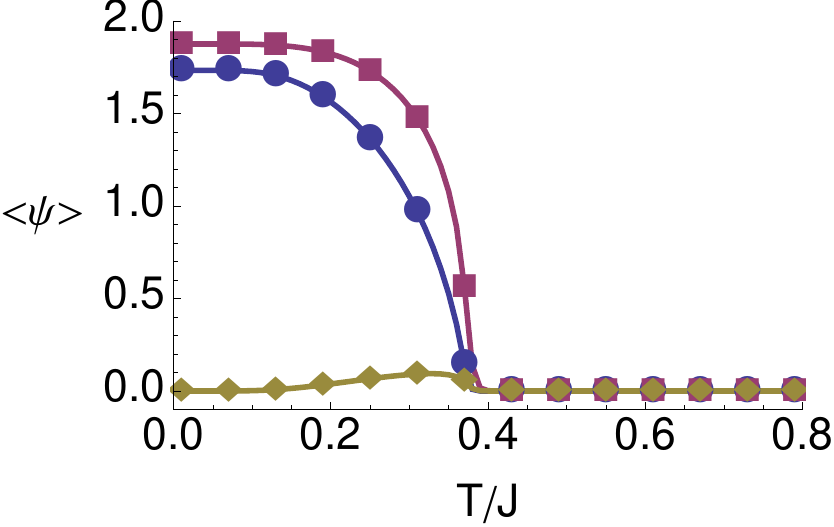}
\caption{(Color online) Temperature dependence of order parameters for
  $J'/J=0.2$, $V/J=0.1$.  For these parameters, there is a direct,
  continuous, normal to AFM transition, at $T/J \approx 0.38$.  The
  three curves show: squares (red online) $|\langle {\bm
    T}_{A,\alpha}-{\bm T}_{B,\alpha}\rangle|/2$, circles (blue online)
  $\langle Q_A^{3z^2}+Q_B^{3z^2}\rangle /2$, and diamonds (yellow
  online) $|\langle {\bm j}_A - {\bm j}_B\rangle |/2$.  Note: in
  Figs.7-10, the symbols are {\sl not} the data points (which are much
  more dense) -- they simply label the different curves.}
\label{fig:ops2010}
\end{figure}

\begin{figure}[htp]
\includegraphics[width=8cm]{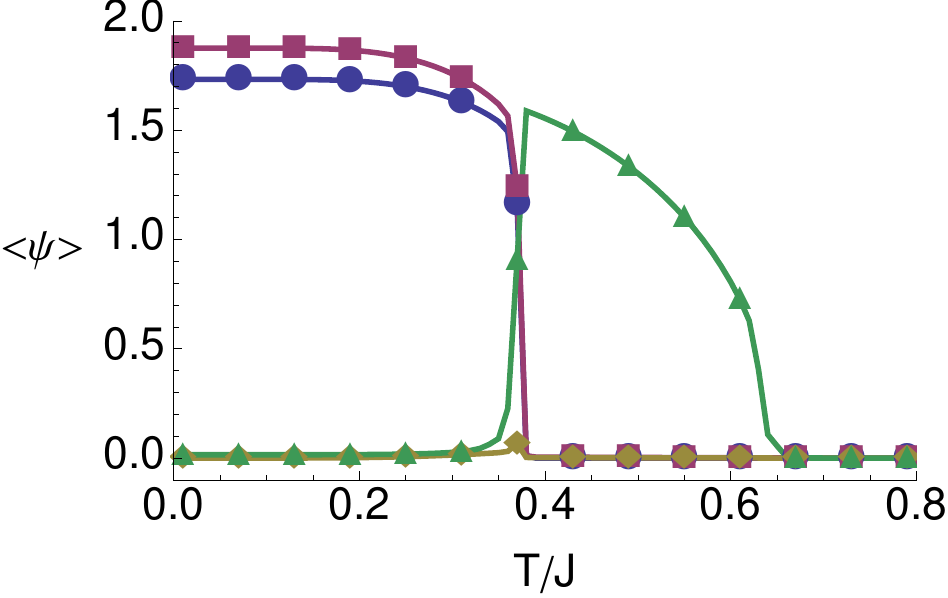}
\caption{(Color online) Temperature dependence of order parameters for
  $J'/J=0.1$, $V/J=0.3$.  For these parameters, there is a continuous
  normal to quadrupolar transition, at $T/J \approx 0.65$, followed by a
  first order transition to the AFM state at $T/J\approx 0.37$.  The
  four order parameters plotted are: squares (red online) $|\langle {\bm T}_{A,\alpha}-{\bm
    T}_{B,\alpha}\rangle|/2$, circles (blue online) $\langle
  Q_A^{3z^2}+Q_B^{3z^2}\rangle /2$, diamonds (yellow online)
  $|\langle {\bm j}_A - {\bm j}_B\rangle |/2$, and triangles (green
  online) $\langle Q_A^{x^2-y^2}-Q_B^{x^2-y^2}\rangle/2$.}
\label{fig:ops1030}
\end{figure}

\begin{figure}[htp]
\includegraphics[width=8cm]{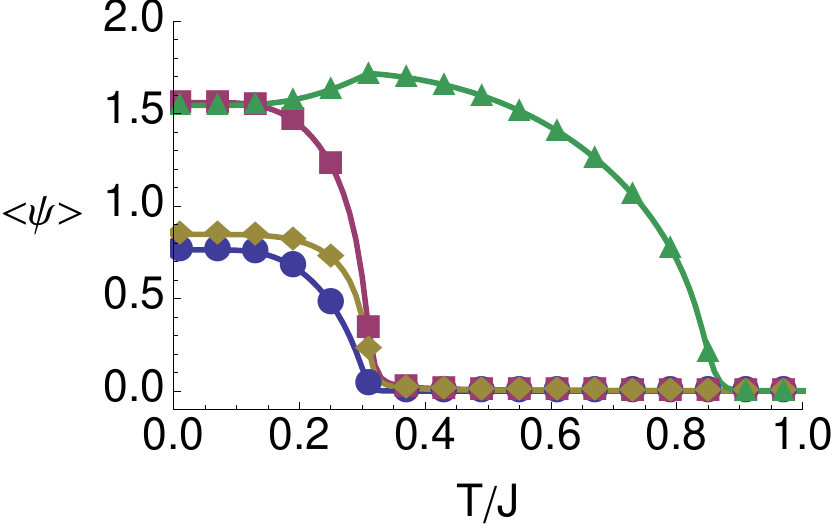}
\caption{(Color online) Temperature dependence of order parameters for
  $J'/J=0.3$, $V/J=0.3$.  For these parameters, there is a continuous
  normal to quadrupolar transition, at $T/J \approx 0.85$, followed by a
  continuous quadrupolar to FM110 transition at $T/J\approx 0.33$.  The
  four order parameters plotted are: squares (red online) $|\langle {\bm T}_{A,\alpha}-{\bm
    T}_{B,\alpha}\rangle|/2$, circles (blue online) $\langle
  Q_A^{3z^2}+Q_B^{3z^2}\rangle /2$, diamonds (yellow online)
  $|\langle {\bm j}_A + {\bm j}_B\rangle |/2$, and triangles (green
  online) $\langle Q_A^{x^2-y^2}-Q_B^{x^2-y^2}\rangle/2$.}
\label{fig:ops3030}
\end{figure}

\begin{figure}[htp]
\includegraphics[width=8cm]{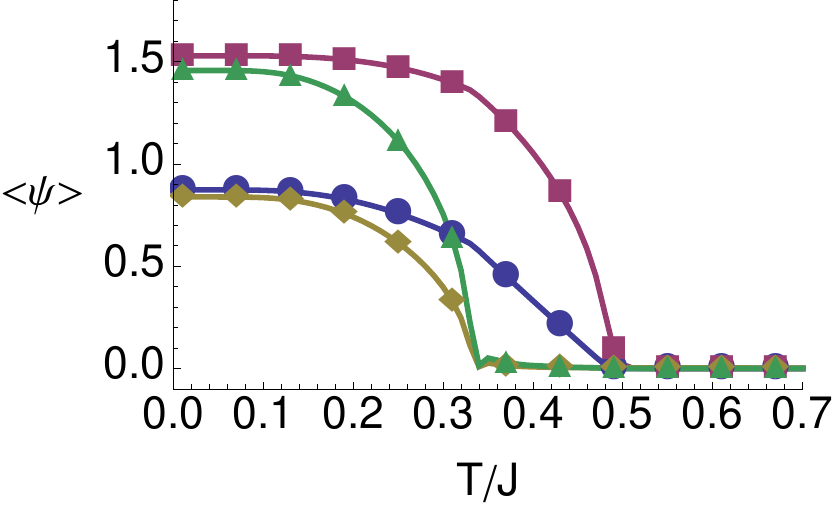}
\caption{(Color online) Temperature dependence of order parameters for
  $J'/J=0.40$, $V/J=0.05$.  For these parameters, there is a continuous
  normal to AFM transition, at $T/J \approx 0.49$, followed by a
  continuous AFM to FM110 transition at $T/J\approx 0.34$.  The
  four order parameters plotted are: squares (red online) $|\langle {\bm T}_{A,\alpha}-{\bm
    T}_{B,\alpha}\rangle|/2$, circles (blue online) $\langle
  Q_A^{3z^2}+Q_B^{3z^2}\rangle /2$, diamonds (yellow online)
  $|\langle {\bm j}_A + {\bm j}_B\rangle |/2$, and triangles (green
  online) $\langle Q_A^{x^2-y^2}-Q_B^{x^2-y^2}\rangle/2$.}
\label{fig:ops4005}
\end{figure}

The transition from the normal to the AFM state is continuous in mean
field theory for small $V$ (see Fig.~\ref{fig:ops2010}), becoming weakly first order for larger
$V$, close to $V_c$ where the intermediate quadrupolar phase
emerges. The normal-AFM transition is characterized, for each of the
three wavevectors, by a two-component primary order parameter, which
could be taken to be the two components of ${\bm T}_{A,\alpha}$ normal
to ${\bm Q}$.  In principle, the degeneracy of the ordering pattern
within this ``XY'' plane normal to ${\bm Q}$ is, as we have remarked,
accidental, and should be removed by additional effects.  We do not,
however, observe this degeneracy lifting within mean field theory for
the present model.  With the degeneracy, the transition should be
therefore described by the free energy for some six component order
parameter.  As we do not understand the degeneracy lifting mechanism
at present, we do not attempt here to establish the true critical
properties for this transition (when it is continuous) with
fluctuations taken into account.

The quadrupolar to FM110 transition is continuous in mean field theory
(see Fig.~\ref{fig:ops3030}). This could be anticipated by examining the
form of the order parameters in the FM110 phase.  We note that the
antiferro-biaxial nematic order parameter of the quadrupolar state is
already non-vanishing in the FM110 phase.  Hence, we might naturally
expect, upon heating, that thermal fluctuations first restore
time-reversal symmetry, yielding the quadrupolar phase, before fully
restoring all symmetry in the normal state.  To determine the nature of
the order parameter for this transition, note that the wavevector and
local anisotropy axes are already established in the quadrupolar state.
Hence the direction of the uniform and staggered magnetizations are
already determined, up to a sign and interchange, above the transition.
For instance, for the quadrupolar state in Eq.~\eqref{eq:11}, with ${\bm
  Q}=2\pi(001)$, the uniform magnetization can lie along $\pm (110)$ and
the staggered magnetization along $\pm(1\overline{1}0)$, or
vice-versa. Thus the symmetry breaking from the quadrupolar to the FM110
state is described by two Ising order parameters.  We therefore expect
this transition, beyond mean field theory, to be similar to that of an
Ashkin-Teller or similar models.

The quadrupolar to AFM transition appears strongly first order (see Fig.~\ref{fig:ops1030}).  This
is in agreement with the expectations of Landau theory, as the
symmetry of the AFM phase is not a subgroup of the symmetry of the
quadrupolar one.  In terms of order parameters, this is evident since $\langle
Q^{x^2-y^2}_i \rangle$ is non-zero in the quadrupolar phase but zero in
the AFM one, while the magnetic order parameters are zero in the quadrupolar
phase but non-zero in the AFM.  Fine tuning of the free energy would
be required to arrange both these types of order to change at the same
temperature in a continuous fashion.  

In region III one encounters a transition from the AFM to FM110 state.
This appears to be continuous in mean field theory (see
Fig.~\ref{fig:ops4005}).  One can understand this by noting that the AFM
solution can be regarded as a subset of the FM110 one, if the unit
vector $(u_1,u_2,0)$ is taken to be along $(1,-1,0)$.  Then the
transition to the FM110 is described by the emergence of a non-zero $m$.
Like the normal to AFM transition, because we have not understood the
degeneracy-breaking mechanism in the AFM state, we do not attempt to
analyze this transition beyond MFT.

\subsection{Effects of anisotropy}
\label{sec:effects-anisotropy}

We now consider the effects of anisotropy on the $T>0$ phase diagram,
focusing on the case of weak $|D| \ll J,J',V$.  We have already
considered the effects of $D$ on the AFM and FM110 states in
Sec.~\ref{sec:transitions}.  We saw that easy-axis anisotropy favors
states with the wavevector ${\bs Q}$ parallel to the $z$ (easy) axis.  This
is because the anisotropy couples directly to the $Q_i^{3z^2}$ field:
\begin{equation}
  \label{eq:16}
  H_D = - D\sum_i (j_i^z)^2 = {\rm const.}  - \frac{D}{\sqrt{3}}
  \sum_i Q_i^{3z^2}.
\end{equation}
Both the AFM and FM110 states have a non-zero and constant
expectation value of $Q_i^{3z^2}$, which is maximized in this
orientation.  Conversely, easy-plane anisotropy favors states with the wavevector
${\bs Q}$ perpendicular to the $z$ (hard) axis, for the same reason.  

\begin{figure}[htp]
\includegraphics[width=8cm]{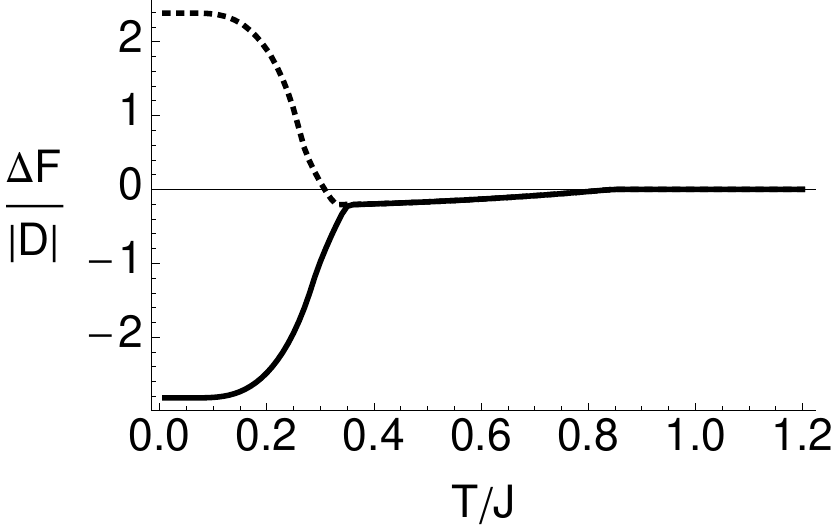}
\caption{Temperature dependence of the free energy difference between
  states with wavevector parallel and perpendicular to the Ising axis,
  in the presence of a weak anisotropy $|D|=0.05$.  Here
  $J'/J=V/J=0.3$.  Solid line: $\Delta F = F({\bm Q}\parallel {\bm\hat
    z};D=.05) - F({\bm Q}\perp {\bm\hat z};D=.05)$.  Dotted line:
  $\Delta F = F({\bm Q}\parallel {\bm\hat z};D=-.05) - F({\bm Q}\perp
  {\bm\hat z};D=-.05)$.  One sees that in the quadrupolar phase, both
  signs of anisotropy favor the wavevector aligned with the z axis.
  In the FM110 phase, however, this is favored only for $D>0$.  For
  $D<0$ (easy plane anisotropy), the state with wavevector normal to z
  is preferred.  Note also that the energy difference is much larger
  in the FM110, consistent with the expected linear and quadratic
  dependence on $D$ in FM110 and quadrupolar phases, respectively.  }
\label{fig:DeltaF}
\end{figure}

We now repeat this analysis for the quadrupolar state.
Here the situation is more subtle because $\langle Q_i^{3z^2}\rangle$
vanishes in the quadrupolar state.  Moreover, the cubic rotations
(e.g. $\langle Q_i^{3x^2}\rangle$), while not vanishing, give zero net
contribution due to the opposite signs on the A and B sublattices.
This means that the splitting of the different wavevector states
vanishes at linear order in $D$.  There is instead a quadratic
contribution, which, numerically, we find favors the states with ${\bm
  Q}$ parallel to $z$ (see Fig.~\ref{fig:DeltaF}).  Being quadratic in $D$, this same
configuration is favored {\em for both the easy-axis and easy-plane
  case}.  Thus we have the interesting situation that for easy-plane
anisotropy, the wavevector orients parallel to $z$ in the
quadrupolar phase, but perpendicular to $z$ in the low
temperature phase.   Note that the quadrupolar phase remains distinct
from the normal phase even with non-zero $D$, as it continues to break
symmetries, notably translational invariance.  

\subsection{Magnetic susceptibility}
\label{sec:field}

In this subsection, we discuss the magnetic response at $T>0$, which is
an important indicator, especially of the quadrupolar ordering
transition.  At high temperature, of course, one observes Curie-Weiss
behavior.  For the general Hamiltonian with anisotropy $D$, there are
two different Curie-Weiss temperatures, for fields parallel and
perpendicular to $z$:
\begin{eqnarray}
  \label{eq:17}
  \Theta_{CW}^{zz} & = &  -\frac{J}{5} + \frac{32 J'}{45} +
  \frac{4D}{5}, \nonumber \\
  \Theta_{CW}^{xx} & = &  -\frac{J}{5} + \frac{32 J'}{45} -
  \frac{2D}{5}.
\end{eqnarray}
These are obtained from the high temperature expansion of the
susceptibility up to $O(1/T^2)$.  These expressions may be useful in
extracting exchange constants from experiment.  Interestingly, if one
calculates the powder average average susceptibility, the contributions
of the anisotropy cancel at this order and the Curie-Weiss temperature
measured in this way is independent of $D$.  It is also interesting to
note that, in the region of larger $V/J$ and small $J'/J$, one obtains a
ferromagnetic ground state with an antiferromagnetic (negative)
Curie-Weiss temperature.  

\begin{figure}
  \centering
  \includegraphics[width=7.5cm]{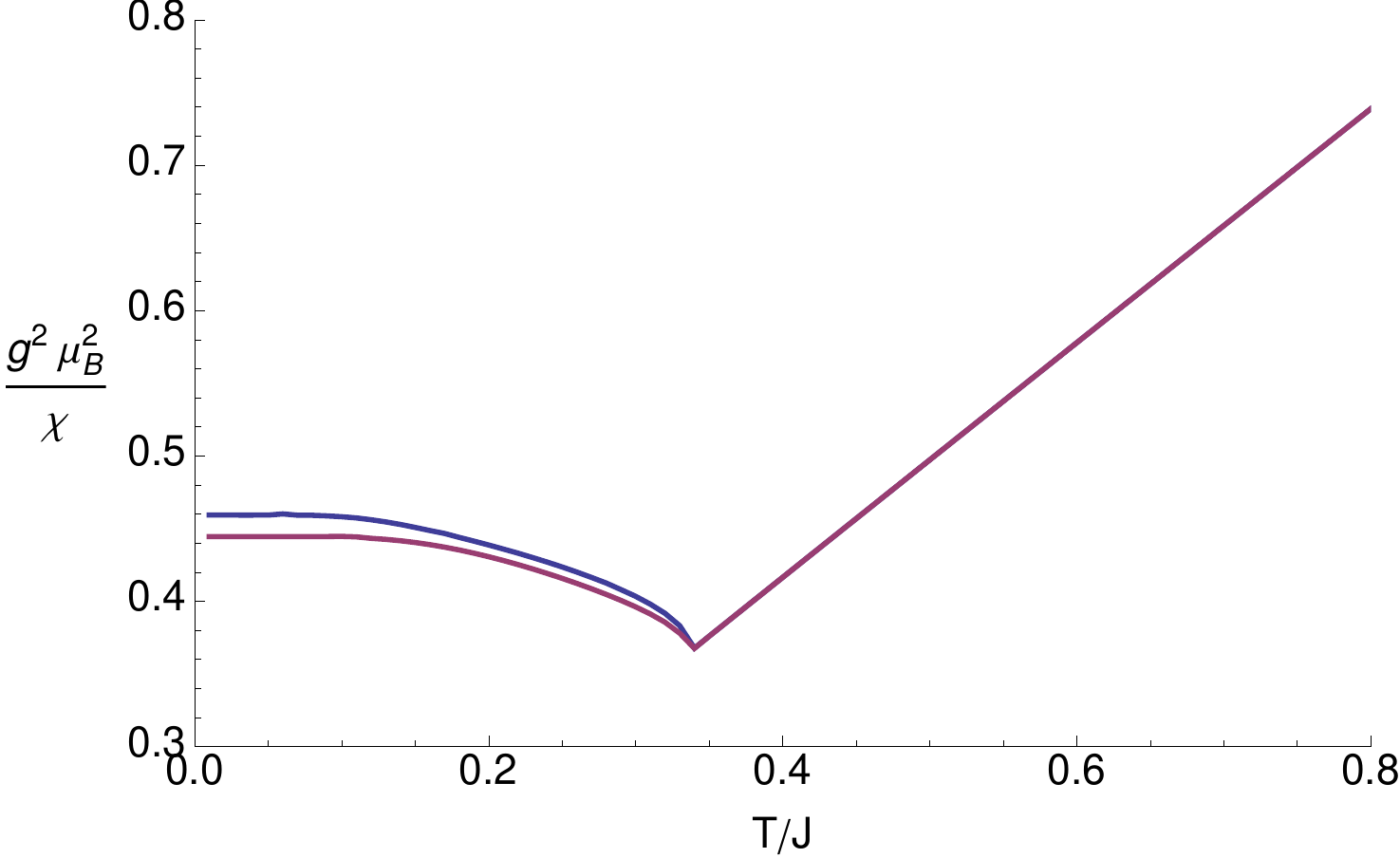}
  \caption{(Color online) Inverse susceptibility at the normal to AFM transition for
    $J'=V=0.1J$.  Blue (upper) curve: $1/\chi_{xx}$, red (lower) curve: $1/\chi_{zz}$. }
  \label{fig:chiinv0101}
\end{figure}

On lowering temperature, the susceptibility shows distinct behaviors in
the different parts of the phase diagram. We focus here for simplicity
on the cubic system, starting with region I.  Here the susceptibility
displays the usual cusp associated with antiferromagnetic order, at the
normal to AFM transition. The inverse susceptibility is plotted in
Fig.~\ref{fig:chiinv0101} for $J'=V=0.1J$, in the midst of region I.  It
shows a minimum at the transition, and pronounced curvature below the
transition temperature, saturating to a large constant value in the
$T\rightarrow 0$ limit. We note that the large zero temperature
susceptibility is not related to gapless excitations, but is a general
consequence of strong SOC, and should be expected in all parameter
regimes of this model.

\begin{figure}
  \centering
  \includegraphics[width=7.5cm]{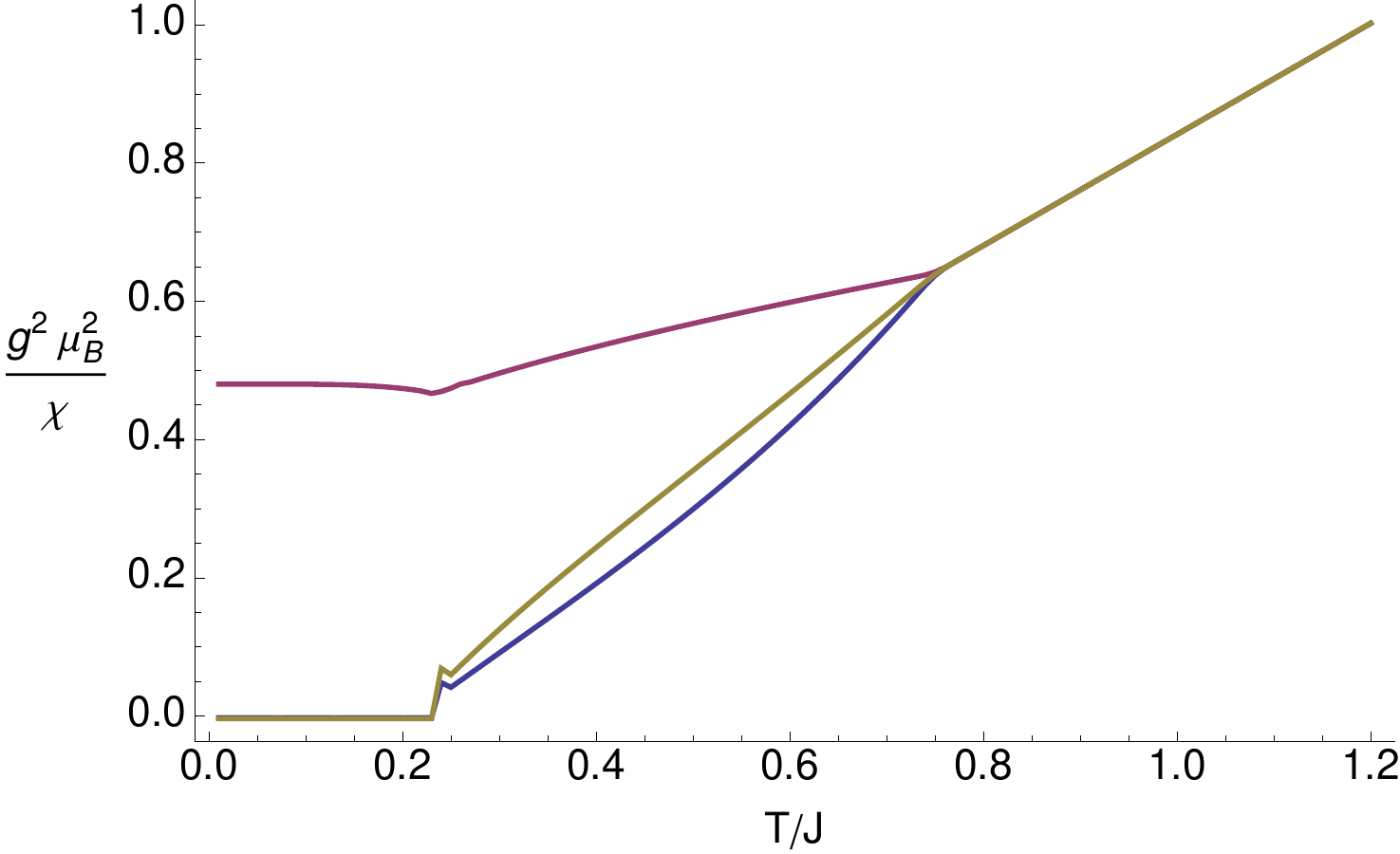}
  \caption{(Color online) Inverse susceptibility for for
    $J'=0.2J$, $V=0.3J$.  Blue (lower) curve: $1/\chi_{xx}$, red (upper) curve:
    $1/\chi_{zz}$, yellow (middle) curve: $1/\chi_{\rm powder}$.  For these
    parameters the quadrupolar transition is at $T/J \approx
  0.75$, and the ferromagnetic transition is at $T/J \approx 0.23$. }
  \label{fig:chiinv0203}
\end{figure}

Next consider region II.  Here, one observes a cusp at the normal to
quadrupolar transition.  This cusp is, however, rather different from
the one just mentioned.  Specifically, it is {\sl not} a minimum of
$1/\chi$, and instead separates two distinct ``Curie-Weiss'' regimes in
which $1/\chi$ is linear but with different, positive, slopes
(i.e. different effective magnetic moments).  The presence of a lower
temperature Curie-Weiss regime is a signature of quadrupolar order.
This is because the quadrupolar mean field splits only the point group
degeneracy of the spins, but preserves a local Kramer's doublet.  This
doublet gives rise to a Curie law.  An example is plotted in
Fig.~\ref{fig:chiinv0203}.  As the quadrupolar order lowers the symmetry
of the system to tetragonal, we see actually two different effective
moments in susceptibility parallel to the wavevector ${\bm Q}$ ($\chi_{zz}$) and
perpendicular to it ($\chi_{xx}=\chi_{yy}$).  We observe that the
effective magnetic moment seen in $\chi_{xx}$ is typically {\sl
  enhanced} in the quadrupolar phase, while it is suppressed in
$\chi_{zz}$, both relative to the isotropic effective magnetic moment in
the normal phase.

At still lower temperature, one encounters the ferromagnetic phases.
Here of course the susceptibility for the easy directions diverges.
Focusing on the dominant FM110 phase, one sees that since the easy
direction is in the (001) plane selected by the quadrupolar order,
$\chi_{zz}$ does not diverge, but $\chi_{xx}$ and $\chi_{yy}$ do.


\section{Beyond mean-field: spin waves and non-magnetic ground states}
\label{sec:sec5}

In Sec.~\ref{sec:sec33}, we obtained the mean field phase diagram.
Here we consider quantum effects beyond mean field.  We first consider
spin wave fluctuations, and obtain the collective mode spectrum in
linear spin wave theory.  From this, we obtain the quantum correction
to the order parameter, and, in the ideal case of $J'=V=0$, we will
see that this is very large and invalidates the mean field theory in
the vicinity of this parameter regime.  This suggests the possibility
of very different states dominated by quantum fluctuations.  We then
explore this possibility, considering some candidate non-magnetic
ground states of our model.

First, we consider the quantum ground states of pairs of sites,
unveiling a pseudo-singlet structure, analogous to the $S=0$ singlet
ground states for pairs of antiferromagnetically interacting spins
with SU(2) symmetry.  This leads naturally to the possibility of
``valence bond'' states built from these pseudo-singlets.  We consider
both a static, Valence Bond Solid (VBS) state, and states in which the
valence bonds are fluctuating, in which case we obtain a Quantum Spin
Liquid (QSL) state.  

Very little theoretical work has been done on QSL states in systems
with strong spin orbit coupling, i.e. with strongly broken SU(2)
symmetry.  As such, the structure of possible QSL states in the
present model requires particular investigation.  Guided by the
pseudo-singlet structure, and the hidden SU(2) symmetry of the model,
we construct candidate QSL states for the full Hamiltonian,
$\tilde{\mc H}_{\text{ex-1}}$, by a slave-particle technique.

\subsection{Spin waves}
\label{sec:sec33}

In the previous two sections, we have discussed the state phase
diagram based on mean-field theory.  Here, we perform a linear spin
wave analysis, which perturbatively describes the effect of quantum
fluctuations on the various phases obtained so far, and also predicts
the structure of collective modes, which might, e.g., be observed in
inelastic neutron scattering.  Finally, because we have not explored
the full space of mean-field states, the calculation also provides an
important check that the phases we have found are at least
metastable.  

The conventional Holstein-Primakoff (HP) transformation for spin-S
operators cannot be directly applied for the three variational ground
states because none of the three states is a fully polarized state for
any projection of the spin angular momentum operator ${\bs j} $. This
is especially severe for the AFM state, for which the spin expectation
value simply vanishes. Instead, we formulate an ``SU(4) spin wave
theory'', by rewriting the Hamiltonian, \eqref{model}, in a bilinear
form in terms of the 15 generators of the SU(4) group.  To do so, we
introduce, for any local basis for the single-site Hilbert space $\{
|n\rangle\},n=1,2,3,4$, the complete set of operators\cite{PhysRevLett.81.3527}
\begin{equation}
  \label{eq:2}
  \mc S^n_m = |m\rangle \langle n|.
\end{equation}
These SU(4) generators obey the algebra $[\mc S^n_m,\mc
S^l_k]=\delta_{nk}\mc S^l_m-\delta_{ml}\mc S^n_k$. We can then use the
HP transformation for the generators of SU(4).   In this transformation,
one selects a particular state in the four-dimensional basis to be the
vacuum and introduces three bosons associated with excitations to the
three other states.  For the AFM phase, the classical ground state
is given by Eq.~\eqref{eq:state_aniz2}.  Hence we take, on the A
sublattice ($z_i$ integer), the basis
\begin{eqnarray}
|1\rangle_A &=&\frac{1}{\sqrt2}(|3/2\rangle+e^{i\phi}|-3/2\rangle),\quad|2\rangle_A = |1/2\rangle,\nonumber\\
|3\rangle_A &=&|-1/2\rangle,\quad|4\rangle_A=\frac{1}{\sqrt2}(|3/2\rangle-e^{i\phi}|-3/2\rangle),
\end{eqnarray}
while for sublattice B ($z_i$ half integer),
\begin{eqnarray}
|1\rangle_B &=&\frac{1}{\sqrt2}(|3/2\rangle-e^{i\phi}|-3/2\rangle),\quad|2\rangle_B = |-1/2\rangle,\nonumber\\
|3\rangle_B &=&|1/2\rangle, \quad|4\rangle_B = \frac{1}{\sqrt2}(|3/2\rangle+e^{i\phi}|-3/2\rangle).
\end{eqnarray}
The Hamiltonian $\tilde{\mc{H}}$ in Eq.~\eqref{model} in this basis has a
quadratic form,
\begin{equation}
\tilde{\mc{H}}=\sum_{\langle ij\rangle} C_{klmn}(i,j)\,{\mc S}^l_k(i)\,{\mc S}^n_m(j),\label{HSS}
\end{equation}
where the coefficients $C_{klmn}(i,j)$ (which are straightforward to
obtain, so we do not give them explicitly) depend linearly on $J$,
$J'$ and $V$.

To introduce the HP transformation on sublattice A (B), we choose
$|1\rangle$ as the vacuum which is annihilated by three ``magnon'' 
annihilation operators $a_n$ ($b_n$), $n=2,3,4$. One can think these
three bosons as descending from mixed spin and orbitals fluctuations of
the Hamiltonian before the ${\mc P}_{\frac{3}{2}}$ projection. The HP
transformation is defined as\cite{PhysRevB.60.6584}, for $i$ in the A
sublattice,
\begin{eqnarray}
{\mc S}^1_1(i)&=&M-\sum_{n\neq1} a^\dagger_n(i)a^{\phantom\dagger}_n(i),    \\
{\mc S}^1_n(i)&=&a^\dagger_n(i)\sqrt{ M-\sum_{l\neq1} a^\dagger_l(i) a^{\phantom\dagger}_l(i)},\qquad (n\neq1)\\
{\mc S}^l_n(i)&=&a^\dagger_n(i)a^{\phantom\dagger}_l(i). \qquad\qquad\qquad\qquad \,\,\,(l,n\neq 1)  ,
\label{HPSU4}
\end{eqnarray}
while for $i$ in the B sublattice, the same formula holds with
$a_n(i)$ replaced by $b_n(i)$.  In the above equations $M$ is defined
as the number of columns in the Young tableaux for the representation of
SU(4).  In our case (fundamental representation), we must set $M=1$.
In the generalization to arbitrary $M$, the classical limit where the
classical ground state becomes exact is $M\to\infty$.  However, we
apply this HP transformation directly for $M=1$.  Inserting this into
\eqref{HSS}, we expand it to obtain a quadratic form in the bosonic
operators.   The constant term in the expansion gives the classical ground state energy,
\begin{equation}
\frac{E_{\text{AFM}}}{\mc{N}}=M^2\left(-\frac{J}{2}+J'+\frac{11V}{12}\right),
\end{equation}
independent of the phase $\phi$.  The quadratic terms lead to quantum
corrections. Defining the Fourier transform of the bosonic operators,
the spin-wave Hamiltonian can be organized in the form
$\sum_\mathbf{k}H_\mathbf{k}$ with
\begin{equation}
H_\mathbf{k}=(\;\mc{A}^\dagger_\mathbf{k}\quad\mc{A}^{\phantom\dagger}_\mathbf{-k}\;)\left(\begin{array}{cc}F_\mathbf{k}&G^\dagger_\mathbf{k}\\G_\mathbf{k}&F_\mathbf{k}\end{array}\right)\left(\begin{array}{c}\mc{A}^{\phantom\dagger}_\mathbf{k}\\ \mc{A}^{\dagger}_\mathbf{-k}\end{array}\right)
\end{equation}
where $\mc{A}_{\mathbf{k}}=(a_{2\mathbf{k}},a_{3\mathbf{k}},a_{4\mathbf{k}},b_{2\mathbf{k}},b_{3\mathbf{k}},b_{4\mathbf{k}})$ is the vector of magnon annihilation operators and $F_\mathbf{k}$ and $G_\mathbf{k}$ are $6\times6$ matrices. This spin-wave Hamiltonian can be diagonalized by standard methods.\cite{PhysRevB.76.064418}

\begin{figure}[htp]
\includegraphics[width=6.0cm]{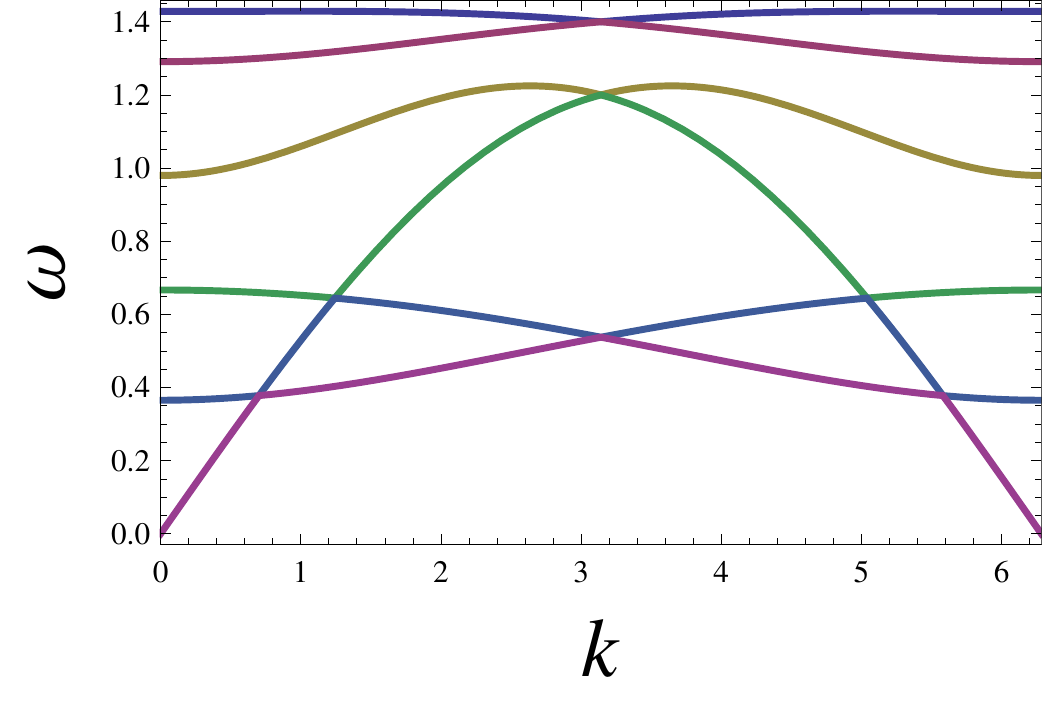}
\caption{Spin wave spectrum for the AFM phase at $J'=0.1$ and $V=0.2$ along [001] momentum direction.  And $J=1$. There is one low-lying gapless mode. The fcc lattice constant is set to be $a=1$. And the phase $\phi =0$ for the ground state in Eq.~\eqref{eq:state_aniz2}. }
\label{fig:spinwave_AFM}
\end{figure}

\begin{figure}[htp]
\includegraphics[width=6.0cm]{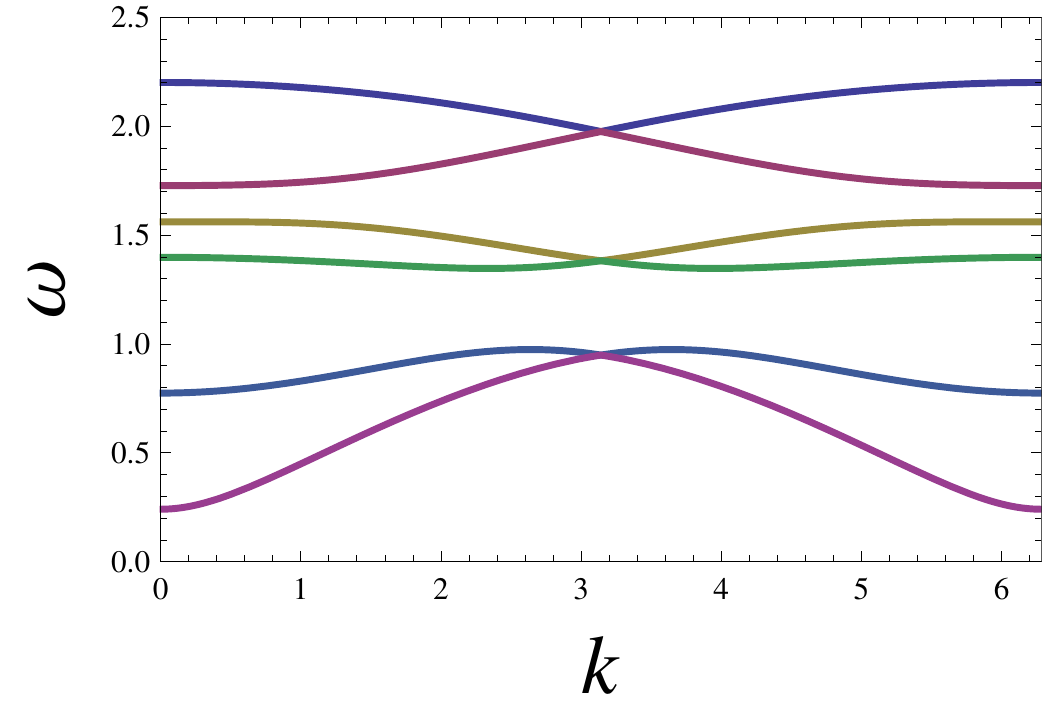}
\caption{Spin wave spectrum for the FM110 phase at $J'=0.3$ and $V=0.2$ along [001] momentum direction. The lowest excitation mode has an energy gap $\Delta = 0.241$ at $k=0$. In the graph, $J=1$. }
\label{fig:spinwave_FM110}
\end{figure}

\begin{figure}[htp]
\includegraphics[width=6.0cm]{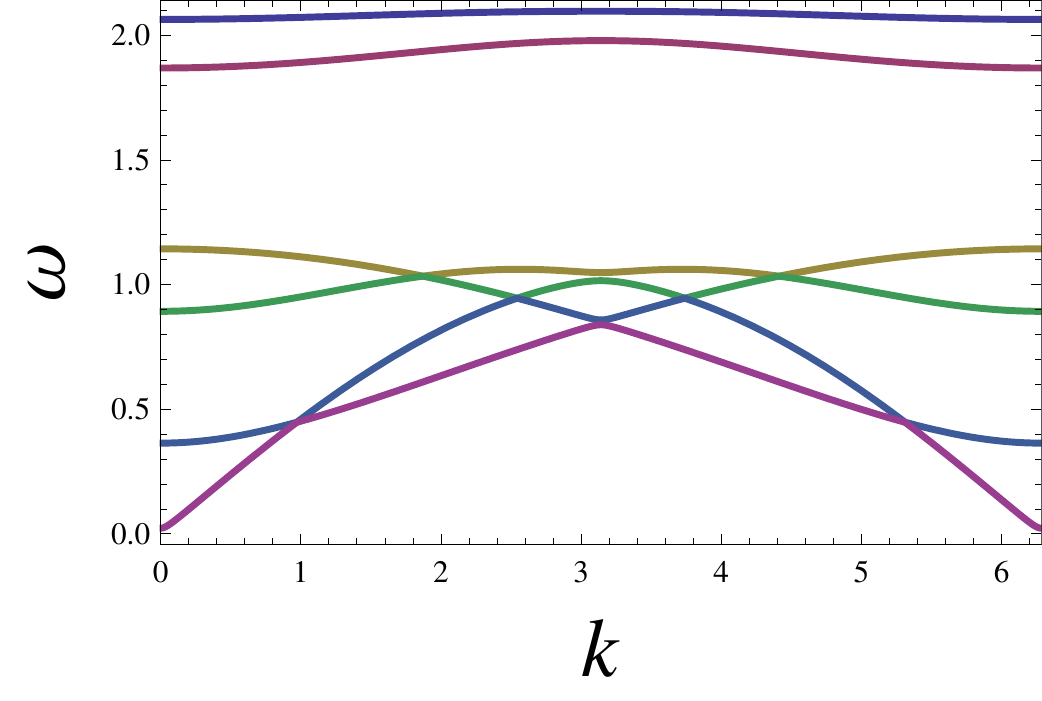}
\caption{Spin wave spectrum for the FM100 phase at $J'=0.1$ and $V=0.4$ along [001] momentum direction. The lowest excitation mode has an energy gap $\Delta = 0.0224$ at $k=0$. In the graph, $J=1$. }
\label{fig:spinwave_FM100}
\end{figure}

For the AFM ground state, Eq.~\eqref{eq:state_aniz2}, we obtain a
gapless ``magnon'' mode, as depicted in
Fig.~\ref{fig:spinwave_AFM}. This gapless mode is associated with the
continuous accidental degeneracy, and indeed occurs for arbitrary
$\phi$.  By contrast, in the FM110 and FM100 phases, one observes a
gap for all the spin wave modes (see Fig.~\ref{fig:spinwave_FM110} and
Fig.~\ref{fig:spinwave_FM100}).   The gap in the FM110 phase increases
with $J'$, as expected since this corresponds to increasingly violated
SU(2) symmetry.  In all cases, the modes are all well-defined with
positive real frequencies, indicating the stability of the phases in
the classical sense: i.e. that we have properly found local energy
minima of the mean field theory.

Finally, having obtained the spin wave modes, we can evaluate the
quantum corrections.  It is most interesting to consider the reduction
of the order parameter by quantum fluctuations.  We can define this by
considering the probability to find a given spin in its mean-field ground
state.  This is nothing but the vacuum state of the HP bosons.  Hence
this probability is given, for a site on the A sublattice, by
\begin{equation}
  \label{eq:21}
  P_{gs}(i)  = \langle 1 - \sum_{n\neq 1} a_n^\dagger(i) a_n^{\vphantom\dagger}(i) \rangle.
\end{equation}
This quantity is directly analogous to the staggered magnetization in
the usual HP treatment of a quantum antiferromagnet.  We therefore
denote $\Delta M = 1-P_{gs}(i) =  \sum_{n\neq 1} \langle a_n^\dagger(i)
a_n^{\vphantom\dagger}(i)\rangle$.  This is
obtained, at $T=0$, by integrating the zero point contribution to the
boson number from each spin wave mode.   
The quadratic spin-wave Hamiltonian is diagonalized
by a Bogoliubov transformation $ \bs{\mc Q}_{\bf k}$,
\begin{equation}
({\mc C}_{\bf k}, {\mc C}^{\dagger}_{- \bf k} )^{T}=  \bs{\mc Q}_{\bf k} ({\mc A}_{\bf k}, {\mc A}^{\dagger}_{- \bf k} )^{T}
\;,
\end{equation}
in which, ${\mc C}_{\bf k} = (c_{1{\bf k}},c_{2{\bf k}},c_{3{\bf k}},c_{4{\bf k}},c_{5{\bf k}},c_{6{\bf k}})$,
and $ \bs{\mc Q}_{\bf k}$ is a $12 \times 12$ matrix.
From this we obtain the quantum correction
\begin{eqnarray}
\Delta M &=& \frac{1}{\mc N} \sum_{n \neq 1}[\sum_{i \in A}  a_n^{\dagger} (i) a_n(i) + \sum_{i \in B} b_n^{\dagger} (i) b_n(i)] \nonumber \\
               &=& \frac{1}{2}\{ \frac{1}{ \mc N }\sum_{\bf k} \sum_{i=1}^{6} [ \bs{\mc Q}^{\dagger}  \bs{\mc Q} ]_{ii}-3 \}.
\end{eqnarray}

Numerically, we find that this quantum correction is maximal for
$J'=V=0$, and is given by $\Delta M \approx 1.7$ at this point. This is
much larger than $1$, implying that the fluctuations at this point are
large and that the mean field theory is at least quantitatively invalid.
For increasing $J'$ and $V$ the correction becomes significantly
smaller, and mean field theory may be reliable.  In the vicinity of the
$J'=V=0$, one may expect a very different ground state, incorporating
strong quantum fluctuations.  We explore some possible {\sl
  non-magnetic} ground states in the remainder of this section.

\subsection{Pseudo-singlets in different planes}

We start our analysis of non-magnetic states by considering two sites
in the XY plane, which interact with the Hamiltonian $\tilde{\mc
  H}_{\text{ex-1}}^{\text{XY}}(i,j)$. Remarkably, the ground state has
a form identical to an SU(2) spin singlet, if written in terms of
pseudospin-$1/2$ states $j^z = \pm 1/2$:
\begin{equation} 
\label{eq:singletxy}
|\text{XY}\rangle_{ij} = \frac{1}{\sqrt{2}} \left(|\tfrac{1}{2} \rangle_i |- \tfrac{1}{2} \rangle_j - |-\tfrac{1}{2} \rangle_i |\tfrac{1}{2} \rangle_j\right)                                     \;.
\end{equation}
One may understand this result by writing down the projected spin and
occupation number operators in $xy$ orbitals, in the basis of $j^z$ eigenstates (see
Eq.~\ref{eq:projxy}):
\begin{eqnarray}
 \tilde{\bs S}_{xy}& = &
\frac{1}{3}
\left( \begin{array}{c|c|c}
0 &  &  \\
\hline
 & \bm{\sigma} &  \\
 \hline
 &  & 0 \end{array} \right),\label{Sxymatrix}
 \\
 \tilde{n}_{xy}  &=& 
\frac{2}{3} 
\left( \begin{array}{c|c|c}
0 &  &  \\
\hline
 & I_2 &  \\
 \hline
 &  & 0 \end{array} \right)\label{Ixymatrix}
 \;.
\end{eqnarray}  
Here $\bm{\sigma}$ is the vector of Pauli matrices and $I_2$ is the $2 \times 2$ identity matrix. One may consider $ \tilde{\bs S}_{xy}$ as an effective spin-$1/2$ operator in the subspace of $j^z = \pm 1/2$ states, which naturally explains the SU$(2)$ singlet in Eq.~\eqref{eq:singletxy}.

For the XZ and YZ planes, one simply needs to apply a cubic
permutation to the results obtained for XY planes, or more formally,
apply a unitary transformation that rotates about the $[111 ]$ axis by
$\pm 2\pi/3$,
  \begin{eqnarray}
    \tilde{S}^{\mu'}_{xz} & = & U^\dagger \tilde{S}^\mu_{xy} U, \label{rotatexz}\\
    \tilde{S}^{\mu''}_{yz} & = & U \tilde{ S}^\mu_{xy} U^\dagger \label{rotateyz}
    \;,
\end{eqnarray}
with
\begin{equation}
  U=\exp\left(-i\frac{2\pi}{3}\frac{j^x+j^y+j^z}{\sqrt{3}}\right).\label{C3rotation}
\end{equation}
The upper indices $\mu'=p(\mu)$ and $\mu''=p^{-1}(\mu)$ in
Eqs.~\eqref{rotatexz} and \eqref{rotateyz} denote cyclic and
anti-cyclic permutations of $x,y,z$, respectively [i.e. $p:(x,y,z)\to
(y,z,x)$, with inverse $p^{-1}:(x,y,z)\to (z,x,y)$]. The two-site
ground states in the XZ and YZ planes are the pseudo-singlets in the
subspace of $j^y = \pm 1/2$ states and $j^x = \pm 1/2$ states,
respectively.

\subsection{Valence bond solid state}
\label{sec:valence-bond-solid}

\begin{figure}
  \centering
  \includegraphics[width=6.0cm]{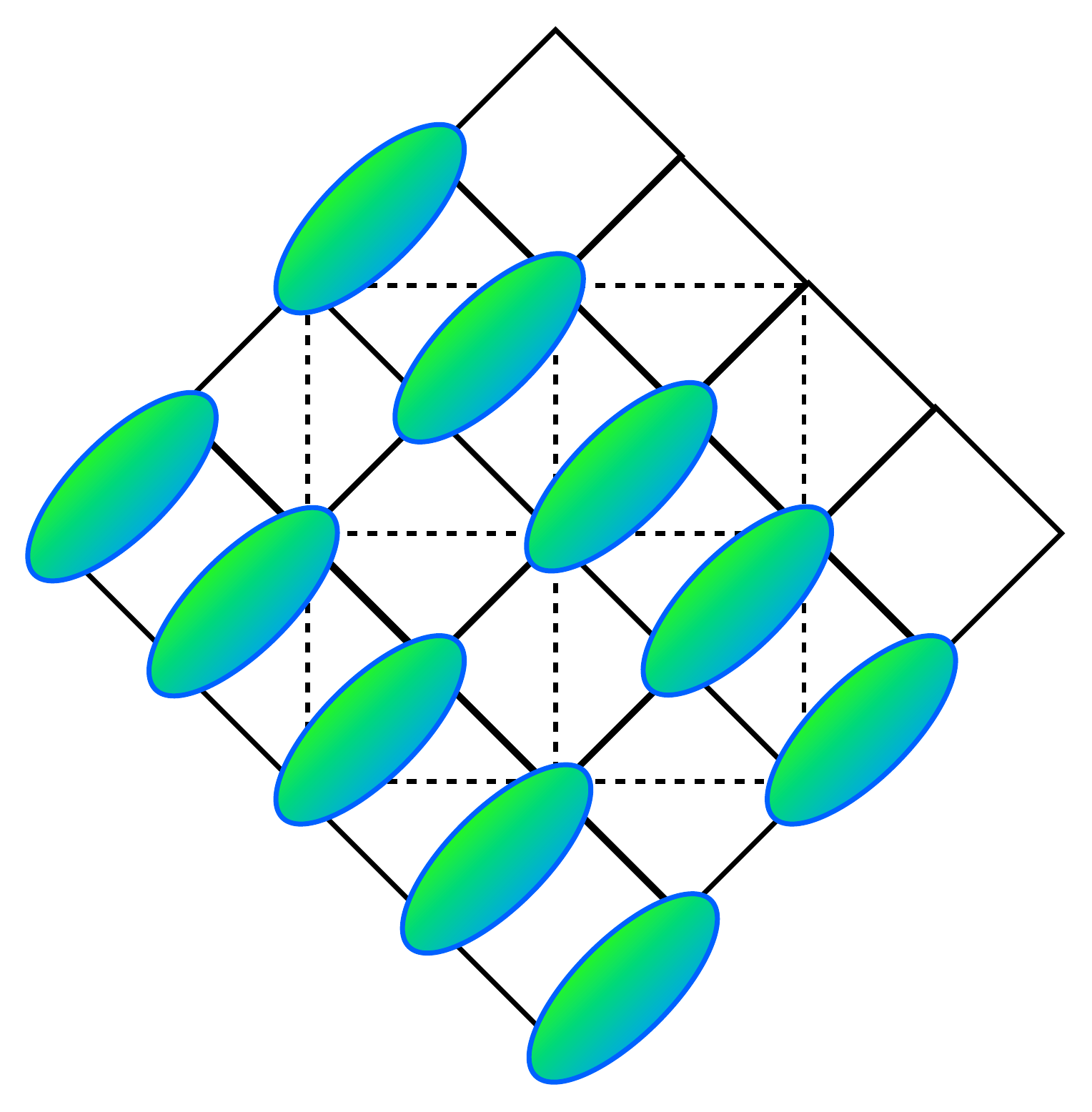}  
  \caption{(color online) Columnar Valence Bond Solid (VBS) state
    within an XY plane.  The dashed square indicates the face of a
    conventional cubic unit cell, while the solid lines connect the
    FCC nearest neighbors within the plane, which form a 45$^\circ$
    rotated square lattice.}
  \label{fig:columnar}
\end{figure}

It is natural to consider a product state of such pseudo-singlet
``valence bonds'' (also called ``dimers'') as a candidate
(prototypical variational) non-magnetic ground state.  To do so, we
must divide the spins into two neighboring sublattices, which will be
paired.  This by necessity breaks lattice symmetries.  Such a state is
called a Valence Bond Solid, or VBS, state.  At the level of valence
bond product states, many possible arrangements of the dimers are
degenerate.  This degeneracy is artificial and will be broken if the
wavefunctions are improved.  We will not investigate this in any
detail, and just consider the simplest VBS state in which the dimers
form a ``columnar'' arrangement within a single (001) plane.  See
Fig.~\ref{fig:columnar}.  

The variational energy of such a state (actually any state with a
planar arrangement of dimers has the same energy) is readily
evaluated.  We obtain the energy per site $E_{VBS} /{\mc N}= \langle VBS|
\tilde{H}_{ex-1}|VBS\rangle/{\mc N} = - 5/12 J =-0.42J$.  This is slightly
higher than the mean-field ground state energy of the AFM state,
$E_{AFM}^{MF}/{\mc N} = - J/2$.  However, the large quantum fluctuations are
expected to destabilize the latter state, and perhaps might stabilize
the VBS one. So such a VBS state seems competitive, and may be
considered as a possibility for future exploration.

\subsection{QSLs and Fermionic mean field theory}

The most general approach that has been applied to describe QSL states
is the slave particle method, in which auxiliary fermions are
introduced, and the ground state for the spin system is described by
some projection of a nominally simple fermionic state into the
physical spin Hilbert space.  This results, in the usual
SU(2)-invariant case, in wavefunctions which are composed of
superpositions of products of SU(2) singlets.  Here, the appearance of
two-site pseudo-singlet ground states points to the possibility of
applying a similar fermionic mean field theory. In this section, we
implement this technique for the full antiferromagnetic exchange
interaction.

We first introduce the auxiliary fermionic creation operators, whose
quanta we call ``spinons''\cite{PhysRevLett.81.3527}
\begin{equation}
|\alpha\rangle_i=f^\dagger_{i\alpha}|\textrm{vacuum}\rangle, \quad\alpha=1,...4,
\end{equation} 
where for convenience we have relabeled the states
$j^z=\frac{3}{2},\frac{1}{2},-\frac{1}{2},-\frac{3}{2}$ by
$\alpha=1,2,3,4$, respectively.  The physical Hilbert space is
constructed from states with one fermion at each site, which imposes
the constraint
\begin{equation}
\sum_{\alpha=1}^4f^\dagger_{i\alpha}f^{\phantom\dagger}_{i\alpha}=1.
\label{constr}
\end{equation}
In this notation, the spin and number operators become
\begin{eqnarray}
\tilde{\bm S}_{i,xy} & \Rightarrow & F^\dagger_i \tilde{\bm S}_{i,xy} F^{\phantom\dagger}_i, \\
\tilde{ n}_{i,xy}    & \Rightarrow & F^\dagger_i \tilde{n}_{i,xy}     F^{\phantom\dagger}_i, 
\end{eqnarray}
where on the right-hand side it is to be understood that the matrices in Eqs.~\eqref{Sxymatrix} and \eqref{Ixymatrix} act on the vector of spinon operators
\begin{equation}
F_i=(f_{i1},f_{i2},f_{i3},f_{i4})^{\textrm{T}}.
\end{equation}
Similar expressions can readily be written for operators in XZ and YZ planes. Thus the Hamiltonian in terms of spinons reads
\begin{eqnarray}
\tilde{\mc H}_{\text{ex-1}}^{\text{XY}} (ij) &=& J  
[(F_i^{\dagger}  \tilde{\bm S}_{i,xy}  F_i)
\cdot (F_j^{\dagger}  \tilde{\bm S}_{j,xy}  F_j)
 \nonumber\\
&&
- \frac{1}{4} 
(F_i^{\dagger}  \tilde{n}_{i,xy}  F_i)
\cdot (F_j^{\dagger}  \tilde{n}_{j,xy}  F_j)
] \nonumber \\
&=& 
\frac{2J}{9} \sum_{\alpha,\beta=2,3}
\left[
  -f^{\dagger}_{i,\alpha} f_{j,\alpha} f^{\dagger}_{j,\beta} f_{i,\beta}
  \right.
  \nonumber \\
  && \left. + f^{\dagger}_{i,\alpha} f_{j,\beta} f^{\dagger}_{j,\beta} f_{i,\alpha} 
-   f^{\dagger}_{i,\alpha} f_{i,\beta} \delta_{\alpha \beta}
\right]
\;,
\end{eqnarray}
in which, $\bm{\sigma}_{\alpha\beta} \cdot \bm{\sigma}_{\alpha'\beta'}
= 2 \delta_{\alpha \beta'} \delta_{\alpha'\beta} - \delta_{\alpha
  \beta} \delta_{\alpha'\beta'}$ has been used.  Similar spinon
Hamiltonians can also be written down for XZ and YZ planes.  When we
write down the full antiferromagnetic exchange Hamiltonian and sum
over XY, YZ and XZ planes, we find that the single-site terms, which
are quadratic in spinon operators, sum up to a constant once we impose
the single occupancy constraint. We are then left with the terms that
are quartic in spinon operators.

We now follow the standard procedure of slave particle mean field theory
to decouple the quartic terms in the spinon Hamiltonian and write down a
mean field ansatz. We start with the exchange Hamiltonian in the XY
plane, $\tilde{\mc H}_{\text{ex-1}}^{\text{XY}}$.

We require a mean field ansatz for the fermionic bond expectation
values, $\langle
f^\dagger_{i\alpha}f^{\phantom\dagger}_{j\beta}\rangle$.  Noting the
structure of the two site pseudo-singlet in this plane, we choose an
ansatz which reproduces a quantum ground state of this type.
Specifically, 
\begin{equation}
\chi_{ij;\alpha\beta}\equiv \langle  f^\dagger_{i\alpha}f^{\phantom\dagger}_{j\beta}\rangle=\chi_{ij}(\mathcal{I}_{xy})_{\beta\alpha},\qquad \langle ij\rangle \in \text{XY}
\label{chiijansatz}
\end{equation} 
with 
\begin{equation} 
\mathcal{I}_{xy}=
\left( \begin{array}{c|c|c}
0 &  &  \\
\hline
 & I_2 &  \\
 \hline
 &  & 0 \end{array} \right)
 \;.
\end{equation}
Note that, by construction, this expectation value is invariant under
the hidden SU(2) symmetry.  The $\chi_{ij}$ on the XZ and YZ planes are
determined by symmetry
\begin{eqnarray}
\label{eq:ansatzxzyz}
\begin{array}{ll}
\vspace{2mm}
\chi_{ij;\alpha\beta}  \equiv  \langle  f^\dagger_{i\alpha}f_{j\beta}\rangle = \chi_{ij}(\mathcal{I}_{xz})_{\beta\alpha},
&
\langle ij\rangle \in \text{XZ} 
\;,
\\
\chi_{ij;\alpha\beta} \equiv  \langle  f^\dagger_{i\alpha}f_{j\beta}\rangle = \chi_{ij}(\mathcal{I}_{yz})_{\beta\alpha},
&
\langle ij\rangle \in \text{YZ} 
\;,
\end{array}
\end{eqnarray}
and 
\begin{eqnarray} 
\label{eq:idxzyz}
\mathcal{I}_{xz} & = & U^{\dagger} \mathcal{I}_{xy} U \\
\mathcal{I}_{yz} & = & U \mathcal{I}_{xy} U^{\dagger}
\end{eqnarray}
with the unitary transformation introduced in Eq.~\eqref{C3rotation}.

We then arrive at the mean field Hamiltonian
\begin{eqnarray}
{\mc H}_{\text{MF}}^{\text{XY}} 
&=& - \tilde{J}  \sum_{\langle ij \rangle \in \text{XY}} \left[
(\chi_{ij} F^\dagger_{j} \mathcal{I}_{xy} F^{\phantom\dagger}_{i}+h.c.)-2|\chi_{ij}|^2\right]
\nonumber\\
&  & +\sum_i \Lambda_i(F^\dagger_iF^{\phantom\dagger}_i-1)
\label{meanfield}
\end{eqnarray}
with $ \tilde{J} \equiv 2J/9 $. 
Here $\Lambda_i$ are the Lagrange multipliers related to the
single-occupancy constraint in Eq.~\eqref{constr}.
$\mathcal{H}_{\text{MF}}^{\text{XZ}}$ and
$\mathcal{H}_{\text{MF}}^{\text{YZ}} $ can be readily written down using
Eq.~\ref{eq:ansatzxzyz} and Eq.~\eqref{eq:idxzyz}.  

\subsubsection{uniform spin liquid}
\label{sec:uniform-spin-liquid}

As discussed in Sec.~\ref{sec:sec2}, the antiferromagnetic exchange
Hamiltonian has a ``hidden'' global SU$(2)$ symmetry, $[G^{\mu},
\tilde{\mc H}_{\text{ex-1}}]=0$. It is easy to find that the full mean
field Hamiltonian we have here respects this ``hidden'' global SU$(2)$
symmetry.  We seek a quantum spin liquid ground state which does not
break any symmetries of the original Hamiltonian. Translational
invariance imposes $\Lambda_i=\Lambda=$ const. First we consider the
ansatz for a uniform spin liquid
\begin{equation}
  \chi_{ij}=\chi_{ji}=\chi ,
\end{equation}
for $i,j$ nearest neighbors on the fcc lattice. This naturally respects
point group and time reversal symmetries. The Hamiltonian in
Eq.~\eqref{meanfield} is then diagonalized by Fourier transform
\begin{equation}
f_{\alpha}(\mathbf{k})=\sum_je^{-i\mathbf{k}\cdot \mathbf{R}_j}f_{j\alpha}.
\end{equation}
We find
\begin{eqnarray}
\mathcal{H}_{MF}&=&\sum_{\lambda=1,2}\sum_{\mathbf{k}}\epsilon_\lambda(\mathbf{k})[\tilde f^\dagger_{\lambda+}(\mathbf{k})\tilde f^{\phantom\dagger}_{\lambda+}(\mathbf{k})+\tilde f^\dagger_{\lambda-}(\mathbf{k})\tilde f^{\phantom\dagger}_{\lambda-}(\mathbf{k})]\nonumber\\
&&+12{\mc N}\tilde{J}\chi^2,
\end{eqnarray}
where $\lambda=1,2$ label doubly degenerate bands with dispersion
\begin{eqnarray}
&& \epsilon_{1,2}(\mathbf{k})\equiv
\tilde{J}\chi\,\tilde{\epsilon}_{1,2}(\mathbf{k})=
-2\tilde{J}\chi\Big[C_xC_y+C_yC_z+C_zC_x \nonumber \\
&&
\pm\sqrt{C_x^2C_y^2+C_y^2C_z^2+C_z^2C_x^2-C_xC_yC_z(C_x\!+\!C_y\!+\!C_z)}\Big]. \nonumber
\\
&&
\end{eqnarray}
Here $C_x=\cos (k_x/2)$, $C_y=\cos (k_y/2)$ and $C_z=\cos (k_z/2)$. The double degeneracy of the two bands is due to Kramer's degeneracy, since ${\bm j}_i$ is a spin-3/2 operator and the Hamiltonian has time reversal symmetry.  

The ground state wave function at the mean field level is described by a Fermi sea of spinons
\begin{equation}
|\Psi_{MF}\rangle=\prod_{\lambda=1,2}\prod_{\mathbf{k}}\tilde f^\dagger_{\lambda+}(\mathbf{k})\tilde f^\dagger_{\lambda-}(\mathbf{k})|\textrm{vacuum}\rangle,\label{fermisurface}
\end{equation}
for all $\mathbf{k}$ below the Fermi surface.  The mean field ground state energy per site is
\begin{equation}
\frac{E_{MF}}{{\mc N} \tilde{J}}=2\chi\sum_{\lambda=1,2}\int\frac{d^3 k}{(2\pi)^3}\,\theta[\mu-\tilde\epsilon_\lambda(\mathbf{k})] \tilde\epsilon_\lambda(\mathbf{k})+12 \chi^2,\label{EMF}
\end{equation} 
where the integral is over the first Brillouin zone of the fcc lattice and the dimensionless chemical potential $\mu$ is fixed by the quarter filling condition
\begin{equation}
2\sum_{\lambda=1,2 }\int\frac{d^3 k}{(2\pi)^3}\,\theta[\mu-\tilde\epsilon_\lambda(\mathbf{k})] =1.
\end{equation}
The wave function in Eq. (\ref{fermisurface}) must be Gutzwiller-projected into the physical Hilbert space with one spinon per site
\begin{equation}
|\Psi\rangle=\mathcal{P}_{n_i=1}|\Psi_{MF}\rangle.
\end{equation}

Here we simply evaluate the ground state energy at the mean field level. Minimizing Eq. (\ref{EMF}) with respect to the parameter $\chi$, we find
\begin{equation}
  \chi^* =-\frac{1}{12}\sum_{\lambda=1,2 }\int\frac{d^3 k}{(2\pi)^3}\,\theta[\mu-\tilde\epsilon_\lambda(\mathbf{k})] \tilde\epsilon_\lambda(\mathbf{k}),
\end{equation} 
and
\begin{equation}
  \frac{E_{MF}}{\mc N}=-\frac{8}{3}J(\chi^*)^2.
\end{equation}
The spinon density is at quarter filling for $\mu\approx -1.58J$. The
mean field energy for the uniform spin liquid state is then
$E^{(0)}_{MF}/{\mc N}\approx -0.041 J $.

\subsubsection{$\pi$-flux spin liquid}
\label{sec:pi-flux-spin}

\begin{figure}[h]
\includegraphics[width=6.0cm]{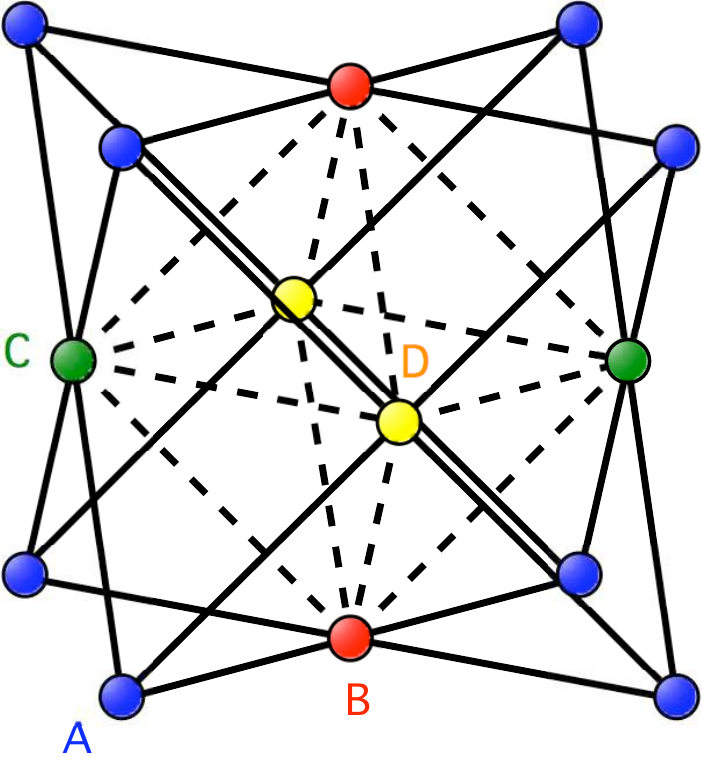}
\caption{(Color online) Conventional unit cell of the fcc lattice divided into four sublattices. The ansatz for the $\pi$ flux state corresponds to assigning hopping amplitude $+\chi$ to the bonds represented by solid lines and $-\chi$ to the bonds represented by the dashed lines. }
\label{fig:piflux}
\end{figure}

We now consider the ansatz for the $\pi$ flux spin liquid state
illustrated in Fig. \ref{fig:piflux}. In order to preserve time reversal
symmetry, the phase of the $\chi_{ij}$ at each bond can only assume the
values $0$ or $\pi$. We divide the fcc lattice into four cubic lattices
\begin{eqnarray}
\mathbf{r}^A_j&=&(0,0,0)+\mathbf{R}_j,\nonumber\\
 \mathbf{r}^B_j&=&(1/2,1/2,0)+\mathbf{R}_j,\nonumber\\
 \mathbf{r}^C_j&=&(0,1/2,1/2)+\mathbf{R}_j,\nonumber\\
 \mathbf{r}^D_j&=&(1/2,0,1/2)+\mathbf{R}_j,\nonumber
\end{eqnarray}
where $\mathbf{R}_j$ is a unit vector in the cubic lattice with lattice
parameter $a=1$. We denote by $F_{jA}$ the vector of spinon annihilation
operators at site $j$ of sublattice $A$, and similarly for the other
sublattices.  We assign $\chi_{ij}=+\chi$ to the bonds connecting sites
in sublattice $A$ to all its nearest neighbors in sublattices $B,C,D$,
and $\chi_{ij}=-\chi$ to the bonds connecting two sites that belong to
sublattices $B,C$ or $D$. In other words, this ansatz corresponds to
assigning $-\chi$ to the three bonds in the $BCD$ base of each
tetrahedron in the fcc lattice and $+\chi$ to the three bonds connecting
the $BCD$ base to the $A$ vertex. As a result, there is $\pi$ flux
through every triangle and zero flux through every square in the fcc
lattice.  While this ansatz is clearly invariant under point group
transformations about $A$ sites, it is also invariant under lattice
translations, despite the fact that this permutes the 4
sublattices. This is because the corresponding changes in $\chi_{ij}$
can be removed by a gauge transformation.  For instance, the gauge
transformation
\begin{eqnarray}
&&F_{jA}\to -F_{jA},\quad F_{jB}\to -F_{jB},\nonumber\\&&F_{jC}\to F_{jC},\quad F_{jD}\to F_{jD},
\end{eqnarray}
exchanges the signs of $\chi_{ij}$ between sublattices $A$ and $B$.  It
follows that the $\pi$ flux ansatz is invariant under point group
symmetries about any site of the fcc lattice and therefore respects all
symmetries of the original Hamiltonian.

Minimizing the energy for the mean field Hamiltonian with four
sublattices, we find 8 doubly degenerate bands for the $\pi$ flux
state. Quarter filling is reached for dimensionless chemical potential
$\mu\approx -1.68J$. The mean field energy is
\begin{equation}
\frac{E_{MF}}{{\mc N}\tilde{J}}=2\chi\sum_{\lambda=1}^8\int\frac{d^3 k}{(2\pi)^3}\,\theta[\mu-\tilde\epsilon_\lambda(\mathbf{k})] \tilde\epsilon_\lambda(\mathbf{k})+12 \chi^2,
\end{equation}
where the integral is over the reduced Brillouin zone of the cubic
lattice. We find $E^{(\pi)}_{MF}/{\mc N}=-0.053J$. This is lower than
the energy for the uniform state.  We may also compare this to the {\sl
  mean field} energy of a VBS state (it is not so meaningful to compare
these slave particle mean field energies directly to the variational energies quoted
earlier for the Weiss mean field and VBS states).   For a mean-field VBS
state, we take $\chi_{ij}$ non-zero only on a set of non-overlapping
dimers.  In this case, we obtain $E_{VBS}^{MF}/{\mc
  N}=-1/18J\approx-0.055J$.  This is slightly lower than the QSL states,
but we expect that the energy of the spin
liquid states will be lowered by the Gutzwiller projection, since the
latter is known to enhance spin-spin
correlations.\cite{citeulike:5750703}

We note that both spin liquid states have Fermi surfaces which are not
nested, and have no obvious instabilities.  The states are also stable
against bond anisotropy which enhances the hopping in a given plane.
Perturbations to NN AFM exchange, such as next-nearest-neighbor
interactions, will in general require more general ans\"atze for the bond
matrix $\chi_{ij}$. Nonetheless, as long as the perturbations are in
some sense small, the $\chi_{ij}$ assumed in Eq. (\ref{chiijansatz}), in
which hopping in a given plane occurs predominantly for two out of four
spinon species, should be a good starting point for approximations.


\section{Discussion}
\label{sec:sec6}


In this paper, we have introduced and analyzed a model to describe
localized electrons in a 4d$^1$ or 5d$^1$ configuration on an fcc
lattice, in which strong spin-orbit coupling and the $t_{2g}$ orbital
degeneracy combine to produce an effective $j=3/2$ description.  The
model contains three interactions -- nearest neighbor antiferromagnetic
and ferromagnetic exchange, and  electric quadrupolar interactions -- and
in addition may
include the effect of structural anisotropy.   We obtain the (Weiss)
mean-field phase diagram, which includes 3 main phases, which all have
a two-sublattice ${\bs Q}= 2\pi (001)$ structure.  In all the phases,
large multipolar order parameters in addition to the usual magnetic
dipolar order are present.  Most remarkably, we find a broad regime of
time-reversal invariant but quadrupolar ordered phase at intermediate
temperatures.  A spin-wave analysis indicates that quantum
fluctuations are strong when nearest-neighbor antiferromagnetic
exchange is dominant, and in this case, we suggest possible quantum
spin liquid and valence bond solid phases. 

\subsection{Experimental ramifications}
\label{sec:exper-ramif}

The theory developed here can be applied and tested in a multitude of
ways.  Here we discuss a few of the main experimental properties which
might be measured.  First, there is the spatial symmetry breaking of the
ordered phases.  All the ordered states, at least in the cubic case,
break lattice symmetries, and in particular double the unit cell.  This
can be tested in experiments such as neutron and x-ray scattering.

We discuss in some further detail the most intriguing case of the
quadrupolar phase, which is non-magnetic.  It constitutes a type of real
(time-reversal invariant) orbital ordering.  It leads to a spontaneous
splitting of the local quadruplet, breaking it down to an elemental
Kramer's doublet.  As such, this is not entirely distinct from a
Jahn-Teller effect, in which ionic motions would lead to such splitting.
In particular, even though in our model atomic displacements are not
involved in an essential way, they would be expected at least to respond
to the orbital order.  In principle, this could be measured by
scattering (x-rays or neutrons) which accurately measure the crystal
structure and symmetry.  From the order parameter description of the
quadrupolar phase, we can obtain the corresponding space group and
crystal structure parameters to be sought in such a measurement.  In
particular, we find that the quadrupolar ordered phase corresponds to
the tetragonal space group P4$_2$/mnm (number 136).  In this space
group, apart from the doubling of the unit cell, one finds that all the
A sites, B sites, and B' sites remain equivalent.  However, the oxygens
are no longer equivalent, but split into three classes, occupying the
4e, 4f, and 4g Wyckoff positions.  Each of these positions has one
degree of freedom which is not fixed by symmetry.  Physically, the
oxygens remain constrained to the simple cubic axes of the perovskite
reference unit cell, but may move by different amounts along each of the
three axes.  This is two more degrees of freedom than is found in the
cubic Fm$\overline{3}$m (number 225) space group, in which the oxygens
maintain an ideal octahedron equidistant from each B (or B') site.
While symmetry requires these structural changes, we do not have at
present an estimate for their magnitude, which could be weak if coupling
to the lattice is not strong.

Another quantity we have already discussed in Sec.~\ref{sec:field}\ is
the magnetic susceptibility, which shows signatures of the quadrupolar
and ordering transitions.  One complication is that the susceptibility
is in many cases highly anisotropic, and one may not be sure what
component(s) are being measured in practice.  Specifically, one may
expect, if the system is ideal and fully in equilibrium, that the broken
symmetry order parameters can be reoriented by the magnetic field, in
such a way that they minimize the free energy.  This will typically
favor orientations which maximize the magnetic susceptibility.  For
instance, in the quadrupolar phase for cubic symmetry, this is an
orientation with ${\bs Q}$ perpendicular to the field.  However, such
reorientation involves motion of domain walls and very large numbers of
spins, and can easily be prevented by pinning or at least be incomplete
in practice.  Thus some diversity of behavior may be expected in
experiment, as well as possibly hysteretic behavior even in the
non-magnetic state.  

When the crystal is non-cubic, one may explore the influence of
single-ion anisotropy on the magnetic susceptibility.  
A na\"ive application of Eq.~\eqref{eq:17} would immediately imply that
single-ion terms do not contribute to the Curie-Weiss temperature as
measured in the powder susceptibility.  However, we caution that these
equations hold only in the true high-temperature regime, in which $T \gg
|D|$.   If $T$ is smaller than or comparable to $|D|$, higher order
terms in the high temperature expansion are non-trivial, and a
non-vanishing fitted Curie-Weiss temperature may result from $D$ alone.
Let us consider the powder susceptibility $\overline{\chi}$ for
independent ions (i.e. neglecting exchange).  One has
\begin{equation}
  \label{eq:18}
  \overline{\chi} = \frac{\chi_{zz}}{3} + \frac{2\chi_{xx}}{3} =
  \frac{3}{4T} + \frac{\tanh{D/T}}{2D},
\end{equation}
in units of $g^2\mu_B^2$.  We then suppose a linear fit to $1/\overline{\chi}$
versus $T$ is made over a narrow region in the neighborhood of the
temperature $T_{\rm fit}$, and extrapolated to find the Curie-Weiss
temperature as the intercept of the horizontal axis.  The result is
\begin{equation}
  \label{eq:19}
  \Theta_{\text{CW}}(T_{\rm fit}) = - \frac{2T_{\rm fit} \left(  T_{\rm
        fit} \sinh (\frac{2D}{T_{\rm fit}})-2 D\right) }{D\left(3
      \cosh(\frac{2D}{T_{\rm fit}})+7\right)}.
\end{equation}
Note that the fitted Curie-Weiss temperature is always negative, and is
independent of the sign of $D$.  It reaches a maximum  in magnitude (at
fixed $D$) of
$\Theta_{\text{CW}} \approx -0.18 |D|$ when $T_{\rm fit} \approx 0.88|D|$, and
only approaches zero very slowly when $T_{\rm fit} \gg |D|$: $\Theta_{\text{CW}} \sim -
\frac{4D^2}{15T_{\rm fit}}$.   Conversely, at a fixed fitting
temperature, the maximum achievable Curie-Weiss temperature is
$\Theta_{\text{CW}} \approx -0.26 T_{\rm fit}$, when $|D|=1.84T_{\rm fit}$.  

\subsection{Materials survey}
\label{sec:materials-survey}

We now turn to a discussion of specific materials which have been
studied experimentally.  

\subsubsection{Ba$_2$YMoO$_6$}
\label{sec:ba_2ymoo_6}

We begin with the material Ba$_2$YMoO$_6$, which has been suggested
experimentally to be an exotic ``valence bond glass'' or to have a
``collective spin singlet'' ground state.  The expected separation
between the $j=3/2$ and $j=1/2$ states in this material is over 2000K,
so that the effective $j=3/2$ description used here should be excellent.
Two recent experimental papers\cite{arxiv0542,arxiv1665}\ observed an
unusual behavior of the magnetic susceptibility, with {\sl two} Curie
regimes, such that $1/\chi$ is linear both above 100K and below 50K.
Moreover, the magnetic specific heat shows a peak around 50K, with
Ref.~\onlinecite{arxiv0542} estimating the total magnetic entropy approximately
equal to R$\ln$4, as expected for $j=3/2$.  Both these results suggest
the existence of some single-ion anisotropy, which would explain the
existence of two Curie regimes because it splits the 4-fold degeneracy
of the $j=3/2$ states but leaves a 2-fold Kramer's doublet at
temperatures below $|D|$, which still gives a Curie signal.  However,
the cubic symmetry observed experimentally seems to rule out such an
explanation.  Moreover, the form of the powder susceptibility in
Refs.~\onlinecite{arxiv0542,arxiv1665} is qualitatively different from that
expected for either fixed easy-plane or easy-axis anisotropy.

These difficulties are resolved if one considers the possibility of
{\sl spontaneous} anisotropy, which indeed is the primary
characteristic of the quadrupolar ordered state.  For example, the
mean-field susceptibility for the cubic model with 
$J'=0.2J, V=0.3J$ is plotted in Fig.~\ref{fig:chiinv0203}.  At temperatures above the
FM110 phase, one indeed observes two Curie regimes in the
susceptibility, with a larger Curie constant at low temperature, as
seen in the experiments.  The kink in $\chi$ coincides with the
quadrupolar ordering transition, and there is a peak in the specific
heat at this temperature, also as observed in experiment.  The
theoretical specific heat has a second peak at lower temperatures,
associated with magnetic ordering and exhaustion of the unsplit
Kramer's doublet.  We suggest that this peak is below the lowest
temperatures measured, or perhaps is avoided due to disorder, and the
spins falling out of equilibrium at low temperature.  The fact that
the Curie-Weiss temperature extracted {\sl below} 50K is only -2.3K
corroborates the notion that any magnetic ordering may be too low to
observe or be obscured by the effects of disorder.

\begin{widetext}

\begin{table}[ht]
\begin{tabular}{|l|c|c|c|c|c|c|c|} 
\hline
Compound & $B'$ config. & crystal structure & $\Theta_{\text{CW}}$ & $\mu_{\text{eff}}(\mu_B)$ & magnetic transition & frustration parameter $f$ & Ref\\ 
\hline \hline 
Ba$_2$YMoO$_6$  & Mo$^{5+}(4d^1)$ & cubic & $-91$K& $1.34$ & PM down to $2$K &  $f \gtrsim 45$ &[\onlinecite{cussen:cm2006}]\\
\hline
Ba$_2$YMoO$_6$  & Mo$^{5+}(4d^1)$ & cubic  & $-160$K& $1.40$ & PM down to $2$K & $f \gtrsim 80$ &[\onlinecite{arxiv0542}]\\
\hline
Ba$_2$YMoO$_6$  & Mo$^{5+}(4d^1)$ & cubic  & $-219$K& $1.72$ & PM down to $2$K & $f \gtrsim 100$ &[\onlinecite{arxiv1665}]\\
\hline
La$_2$LiMoO$_6$ & Mo$^{5+}$ (4d$^1$) & monoclinic & -45K & 1.42 & PM to
2K & $ f\gtrsim 20$ & [\onlinecite{arxiv1665}]\\ 
\hline
Sr$_2$MgReO$_6$ & Re$^{6+}(5d^1)$ & tetragonal & $-426$K & $1.72$  &spin glass, $T_G \sim 50$K & $f \gtrsim 8$ &[\onlinecite{wiebe:prb2003}] \\
\hline
Sr$_2$CaReO$_6$ & Re$^{6+}(5d^1)$ & monoclinic & $-443$K & $1.659$ &spin glass, $T_G\sim 14$K & $f \gtrsim 30$ &[\onlinecite{wiebe:prb2002}] \\
\hline
Ba$_2$CaReO$_6$ & Re$^{6+}(5d^1)$ & cubic to tetragonal (at $T\sim 120$K) & $-38.8$K & $0.744$ & AFM $T_N =15.4$K & $f\sim 2$ & [\onlinecite{yamamura:jssc2006}] \\
\hline
Ba$_2$LiOsO$_6$ & Os$^{7+}(5d^1)$ & cubic & $-40.48$K & $0.733$ &   AFM $T_N\sim8$K    & $f \gtrsim 5$ & [\onlinecite{stitzer:sss2002}]\\
\hline
Ba$_2$NaOsO$_6$ & Os$^{7+}(5d^1)$& cubic & $-32.45$K & $0.677$ &      FM $T_N\sim8$K & $f \gtrsim 4 $ & [\onlinecite{stitzer:sss2002}] \\
\hline
Ba$_2$NaOsO$_6$ & Os$^{7+}(5d^1)$& cubic & $\sim-10$K & $\sim 0.6$ &      FM $T_N=6.8$K & $f \gtrsim 4 $ & [\onlinecite{Erickson}] \\
\hline
\end{tabular}
\caption{A list of ordered double perovskites. Note the discrepancy in Curie-Weiss temperature and $\mu_{ \text{eff} } $ may originate from the experimental fitting of data at different temperature range. }
\label{tab:Tab1}
\end{table}

\end{widetext}

\subsubsection{La$_2$LiMoO$_6$}
\label{sec:la_2limoo_6}

La$_2$LiMoO$_6$ is monoclinic, the deviation from cubic symmetry
arising primarily from rotations of the octahedra.   The local
coordination of the Mo sites is nearly perfectly octahedral with a
weak tetragonal compression.  The nature of crystal field effects, if
significant, is unclear at present.  Magnetically, the susceptibility
shows, like Ba$_2$YMoO$_6$, two apparent Curie regimes, separated by a
kink at approximately 150K.  However, opposite to that material,
La$_2$LiMoO$_6$ shows a smaller effective moment at low temperature
compared to high temperature.  In addition, the high temperature
Curie-Weiss temperature is  $\Theta_{\text{CW}} \approx -45K$, significantly
smaller than the kink temperature.  Irreversibility distinguishing the
behavior of the ZFC/FC susceptibility appears below 25K.  

The appearance of two Curie regimes again suggests either fixed or
spontaneous magnetic anisotropy setting in around 150K.  However, the
{\sl reduction} of the effective moment below the kink in $\chi^{-1}$ is
puzzling.  We did not find this behavior in the powder susceptibility
within our model, with or without anisotropy modeled by $D$.  As
remarked above, however, the actual nature of the crystal field anisotropy
in La$_2$LiMoO$_6$ is unclear.  If it is significant and different in
form from the $D$ term, this might explain the behavior.  Single crystal
studies would be helpful in elucidating the situation.

\subsubsection{Sr$_2$CaReO$_6$ and Sr$_2$MgReO$_6$}
\label{sec:sr_2c-sr_2mgr}

Sr$_2$CaReO$_6$ and Sr$_2$MgReO$_6$ have distorted perovskite
structures, with monoclinic and tetragonal symmetry,
respectively.\cite{wiebe:prb2002,wiebe:prb2003} Experimentally, the
materials are notable for their very high antiferromagnetic
Curie-Weiss temperature, $-\Theta_{\text{CW}}\gtrsim$ 400K.  Susceptibility
and specific heat measurements show anomalies suggestive of freezing
and/or short-range ordering at 14K and 50K, for Sr$_2$CaReO$_6$ and
Sr$_2$MgReO$_6$, respectively.  Two possible interpretations of this
behavior are: (1) the Curie-Weiss temperature is dominated by strong
exchange, but fluctuations largely suppress ordering, or (2) the
Curie-Weiss temperature is due largely to single-ion effects, and the
true exchange scale is comparable to the observed anomalies in $\chi$
and $c_v$.  

In the former scenario, the key question is why these two materials show
so much larger exchange than do the other compounds in this family.
From the point of view of this work, attributing the Curie-Weiss
temperature to exchange alone would imply $J$ is actually comparable to
the SOC, so that the projection to $j=3/2$ may even be suspect.  The
Curie-Weiss temperatures are sufficiently large that one may suspect
that the $5d$ electrons are not so well localized, and the system is
close to a Mott transition.  It would be interesting to measure their
optical properties to address this possibility.

The latter explanation seems possible, as both materials show
significant deviations from the cubic structure: Sr$_2$CaReO$_6$ is
monoclinic, while Sr$_2$MgReO$_6$ is tetragonal.  The actual
distortions of the octahedra are rather small in both cases, the Re-O
distance varying by only about 0.02$\AA$ at room temperature.
However, there are significant rotations and tilts of the octahedra,
and crystal field splittings of the $j=3/2$ quadruplet are certainly
allowed.  Examination of the Re-O bond lengths suggests
easy-axis anisotropy.  From Eq.~\eqref{eq:19}, we see that in principle
a negative Curie-Weiss temperature could be attributed to $D$.
However, from the present model we cannot obtain such a large value,
which in these two materials is comparable or larger than the fitting
temperature.  Nevertheless, we may imagine that some combination of
exchange and single-ion anisotropy may conspire to produce the observed
behavior.  

If we assume a large easy-axis anisotropy, we would then expect, based
on the the analysis in Sec.~\ref{sec:easy-axis-anisotropy}, to have an
AFM ground state.  The anomalies might be related to this ordering.
Experimentally, spin freezing and irreversibility is observed, but
without clear signs of long-range ordering.  The experimentalists
caution that, due to the small magnetic moment of the Re$^{6+}$ ions, a
small ordered component could not be ruled out in either
material.\cite{wiebe:prb2003,wiebe:prb2002} Indeed, in the AFM state, a
very small moment is expected, due to the primacy of octupolar order.

While this is promising, we note that it is likely that several effects
not in our model play a role.  First, the structure of the materials is
not a simple compression of the cubic structure, and so the crystal
fields might have a significantly different form from the simple $D$
term. This is especially true in Sr$_2$CaReO$_6$, which has the more
distorted monoclinic structure.  Second, in Sr$_2$MgReO$_6$, the Re-O-Mg
bond angles are very different in the XY plane (160$^\circ$) and normal
to it (180$^\circ$),\cite{wiebe:prb2003} so substantial spatial
anisotropy in the exchange couplings may be present.  This is not
included in our model.  Also in Sr$_2$MgReO$_6$, the ZFC and FC
susceptibility actually diverge already around 300K, which suggests a
high degree of disorder in this material, which might be responsible for
converting the AFM to a glassy state.

One indication supporting an antiferromagnetic ground state is the
observation, in Sr$_2$CaReO$_6$, of a $T^3$ magnetic contribution to the
specific heat, in contrast to the usual linear one characteristic of a
spin glass.  The $T^3$ behavior would naturally be expected from spin
waves in the AFM state, which as we have noted displays gapless spin
waves, at least in the semi-classical approximation.  Such $T^3$
behavior might even persist if the AFM order had a finite correlation
length, due to Halperin-Saslow modes\cite{halperin1977hts}, as recently
postulated in NiGa$_2$S$_4$\cite{podolsky2008hsm}.  A $T$-linear specific
heat was observed in Sr$_2$MgReO$_6$, albeit with a small
coefficient.\cite{wiebe:prb2003} As we have already remarked, however,
this material is likely to be more disordered, consistent with the more
conventional spin glass-like specific heat.

While these considerations seem reasonable, they are hardly
definitive.  Further studies, particularly on single crystals, would
be most helpful in clarifying the physics of these materials.

\subsubsection{Ba$_2$CaReO$_6$}
\label{sec:ba_2careo_6}

In Ba$_2$CaReO$_6$, there is a structural transition from a high
temperature cubic phase to a low temperature tetragonal one, with a
doubled unit cell, at $T=120\text{K}$\cite{yamamura:jssc2006}.  The
experimentalists have fitted the low temperature structure to the I4/m
space group.  From this fit, they found an elongation along the $c$ (or
$z$) axis, but a slight compression of the ReO$_6$ octahedra. One may
consider two possibilities.  Either this is indeed the correct symmetry,
in which case it must have structural origin not related to the 5d
electrons, {\sl or} this transition in fact coincides with the
quadrupolar ordering described here, which also gives a tetragonal state
with the same unit cell.  The P4$_2$/mnm space group was not considered
in Ref.\onlinecite{yamamura:jssc2006}.

A small, negative Curie-Weiss temperature $\Theta_{\text{CW}} = -$39K was
measured by fitting the susceptibility in the high temperature cubic
phase , indicating that $J'$ should be not too large.  A predominantly
antiferromagnetic ordering transition was observed at $T=$15K, which is
consistent with our expectations in the small $J'$ and $V$ regime
(recall that an antiferromagnetic state is expected both with and
without single ion anisotropy in this parameter range).  It would be
interesting to compare the predicted magnetic structure in the AFM or
AFM' phase with experiment, by carrying out neutron scattering and NMR
measurements.

\subsubsection{Ba$_2$NaOsO$_6$}
\label{sec:ba_2naoso_6}

Ba$_2$NaOsO$_6$,\cite{Erickson} is one of only two examples in this
class in which single crystal experiments have been performed, to our
knowledge.  A transition at $T=$6.8K was found to a ferromagnetic state
with easy axis along a $[ 110 ] $ direction.  Within the experimental
resolution, the material was found to remain cubic down to the lowest
measured temperature.  A fit to Curie-Weiss behavior found a negative
Curie-Weiss temperature, $-13$K$<\Theta_{\text{CW}}<-$10K, depending upon field
orientation.  We note that this is compatible with Eq.~\eqref{eq:17},
and suggests that the material is in the regime of larger $V$ and
smaller $J'$.  In this region, we expect a high temperature quadrupolar
transition, well above the ferromagnetic state.  While no such
transition is observed in the experiment, the magnetic specific heat of
such a transition may be masked by the lattice contribution at higher
temperature, and the signature in susceptibility may be subtle.

Several other indications are in favor of this scenario.  First, the
susceptibility continues to display pronounced anisotropy, favoring the
$[ 110 ] $ direction, up to at least 200K, well above the ferromagnetic
transition.  This would indeed be expected in the quadrupolar ordered
state.  Second, the magnetic specific heat divided by temperature,
integrated over the peak up to about 15K gives a magnetic entropy of
approximately $R\ln 2$, which is only half the expected entropy for the
$j=3/2$ quadruplet.  This entropy may be released over a significantly
higher temperature range, up to the quadrupolar ordering transition.
Third, the observation itself of ferromagnetism with a
$[110]$ easy axis is a marked success of our theory.  This
type of anisotropy is not natural from the standpoint of Landau theory,
within which cubic anisotropy manifests itself at leading order as a
term in the free energy of the form $v[(M^x)^4+(M^y)^4+(M^z)^4]$ (${\bm M}$ is
the magnetization), which, depending upon the sign of $v$, generates a
$[ 111]$ or $[ 100]$ easy axis.  A continuous
Landau transition to the FM110 state is instead made possible by the
fact that the quadrupolar order already breaks the cubic symmetry in the
paramagnetic state.

It would be interesting to further probe the system to establish in more
detail the correspondence (or lack thereof) with our theoretical
predictions.  The predicted tetragonal distortion of the cubic structure
would be a natural quantity to seek in experiment.  This also manifests locally in
the magnetism, since although the net magnetization is aligned with the
$[ 110 ]$ axis, the local spin expectation values are not.  This might be
measurable for instance by a local probe such as NMR.

\subsubsection{Ba$_2$LiOsO$_6$}
\label{sec:ba_2lioso_6}

Ba$_2$LiOsO$_6$ has also been grown in single crystal form.  The
structure was determined to be cubic by x-ray diffraction at room
temperature.\cite{stitzer:sss2002} Aside from this diffraction data,
only bulk magnetic susceptibility results are available.   One observes
a negative Curie-Weiss temperature $\Theta_{\text{CW}} \approx -40K$, and an
apparent antiferromagnetic transition at $T_N \approx 8K$.  This appears
largely consistent with the expected behavior in region I of
Fig.~\ref{fig:thermal_phase}.

\subsection{Comparison and future work}
\label{sec:comp-with-other}

We are aware of only one other theoretical work studying this class of
materials.  Lee and Pickett\cite{lee:epl2007} performed electronic
structure calculations for Ba$_2$NaOsO$_6$ and Ba$_2$LiOsO$_6$,
emphasizing the role of SOC.  We completely agree with the conclusion
that SOC plays a crucial role in the magnetism.  However, the magnetic
structure and phase transitions were not addressed.

Our study is much more comprehensive, and gives a great deal of guidance
both for future theory and experiment.  Many experimental suggestions
have already been made.  In particular verification of the quadrupolar
ordering transition would be especially exciting.  On the theoretical
side, the problem of the effects of quantum fluctuations in the small
$J'$ and $V$ limit remains rather open.  It would be remarkable if a
spin liquid or valence bond solid state could be established for this
highly non-SU(2) symmetric and nominally ``large'' spin $j=3/2$ model.
To do so would require some hard theoretical work applying more
quantitative numerical methods to our model Hamiltonian.  A natural
extension of this work would be to consider the ``higher spin'' analogs
of these materials, with 4d$^2$ or 5d$^2$ electronic states.  As there
is still partial occupation of the $t_{2g}$ orbitals in this case, we
expect SOC again to play a dominant role, and interesting multipolar
physics is likely present.

\acknowledgments

We thank John Greedan, Ian Fisher and Ram Seshadri for useful
discussions.  This work was supported by the DOE through Basic Energy
Sciences grant DE-FG02-08ER46524. LB's research facilities at the KITP
were supported by the National Science Foundation grant NSF PHY-0551164.

\bibliography{ref}

\end{document}